# How UX Practitioners Produce Findings in Usability Testing

STUART REEVES, School of Computer Science, University of Nottingham

Usability testing has long been a core interest of HCI research and forms a key element of industry practice. Yet our knowledge of it harbours striking absences. There are few, if *any* detailed accounts of the contingent, material ways in which usability testing is *actually practiced*. Further, it is rare that industry practitioners' testing work is treated as indigenous and particular (instead subordinated as a 'compromised' version). To service these problems, this paper presents an ethnomethodological study of usability testing practices in a design consultancy. It unpacks how findings are *produced* in and as the work of observers analysing the test as it unfolds between moderators taking participants through relevant tasks. The study nuances conventional views of usability findings as straightforwardly 'there to be found' or 'read off' by competent evaluators. It explores how evaluators / observers collaboratively work to locate relevant troubles in the test's unfolding. However, in the course of doing this work, potential candidate troubles may also routinely be dissipated and effectively 'ignored' in one way or another. The implications of the study suggest refinements to current understandings of usability evaluations, and affirm the value to HCI in studying industry practitioners more deeply.

## 1 INTRODUCTION

Usability testing has long been a core interest of HCI's focus on evaluation methods. Although a significant, rich body of work elaborates it, there are striking absences. Firstly, though various forms of usability testing are routinely employed in industry to shape actual product and service design outcomes, there are few, if any, studies of the most frequent practitioners of usability testing, namely industry-based UX and design professionals. Secondly, there is a paucity of detailed empirical accounts of *how* usability testing and its many variants are concretely brought off as practical lab work, i.e., "the hard work required to get methods to work" [16]. We know little about how the usability findings at the core of this work come to be *produced*, in and as the material work of evaluators. This paper begins filling in both gaps via an empirical investigation into usability evaluation methods in use. Simultaneously, the paper elaborates and demystifies one of the many components of everyday design work.

Conceptually, the bulk of HCI research on the topic typically frames usability problems as intrinsic to artefacts being evaluated: in other words, that usability findings are 'there to be found', to be 'read off' by competent evaluators from the artefact in question. In this view usability findings are *invariant properties* of the object under test (often to be counted in some way, cf. Lindgaard's critique [61]). In contrast to this, I seek to advance the idea that usability findings are concerted methodological achievements of evaluators *seeing trouble*. In this vein I want to posit that findings are interactionally produced: a matter of evaluators working up, and jointly formulating witnessed events as recognisable objects that are relevant to matters of usability and design. As part of this I want to also show how production work is necessarily and irremediably entangled with test procedures, design work practice, and other complex organisational and stakeholder concerns.

Now, none of this is to say that HCI's rich literature on usability evaluation has somehow ignored *problem identification* as a topic (e.g., see [31, 95]). Rather, HCI remains curiously uninterested in the interactional organisation findings and how they come to be dealt with as inherently social objects. Interestingly, this lack of interest in the formulation work at the core of usability evaluation is not just confined to academic research on the topic but also absent in academic accounts of practitioner approaches as well as in practitioner accounts themselves (e.g., see [78]). By focussing on *practitioners* at work, I am interested in describing the methodic particularities of how findings get formulated in concrete ways within lab-based industry settings. That said, it does not seem to be the case that such oversights are completely unknown. Hornbæk has argued several "dogmas" related to problem identification and evaluation processes require a second look [51], while there have also been repeated calls for HCI research to improve how it takes the realities of practice into account [98, 61, 102].

But how might we start to better understand the shape of the absences I describe? In order to get empirical purchase on the production of usability findings as an interactional matter, this paper investigates sites of usability testing (i.e., "usability work"[1] that involves "empirical evaluation" [17] as opposed to "analytical"). It does this via an ethnomethodological study of lab-based usability tests run by practitioners.

Two points of clarification are now required. Firstly, I have repeatedly used the term 'practitioners', which, while frequently used in HCI research literature, is routinely left unexplicated. By practitioners I mean professionals within industry that self-identify their various roles and activities in technology development as (variously) usability engineering, user experience, interaction design, user / UX research, experience design, and user- or human-centred design. Often this language is deployed to affiliate in some way with design. Such industry practitioners are a conglomeration of communities mostly operating outside the boundaries of academic HCI, yet who nonetheless engage in usability testing of one form or another as a vital part of their everyday work. Unfortunately there is a danger that HCI researchers' use of terms like 'practitioner' without such clarification might lead to such communities being overlooked or even trivialised, with key distinctions potentially obscured, including the sometimes conflicting, sometimes complementary roles played by a complex mix of product owners, designers, user researchers, client organisations, etc. This is in spite of indications that HCI *does* indeed have interest in investigating industry design practice itself and the various relationships and frictions between such communities of practice and HCI [22, 16, 42]. In response to this, my study here focusses specifically on how user / UX researchers, product owners, and interaction designers work together to produce findings, and in so doing goes some way to decompress HCI's gloss of 'practitioners'.

The second clarification is to fend off a potential conceptual mistake, namely that practitioners' work is somehow categorically *the same* species of activity as the application of notionally 'the same' techniques in academic research environments. As Wixon points out, this is a non sequitur [98]. In the latter, what is at stake in usability work is dissemination of results to research audiences via formal publication. In the former, usability work turns fundamentally on the role of product owners within organisations as critical stakeholders. At the same time, the suggestive family resemblances of usability evaluation techniques and their application across the differing domains of design communities in industry and HCI research *does* sustain the existence of some connective tissue between the two. Working out the contours of these differences / resemblances remains an ongoing project for HCI [16]; the study I present here can assist with the progress of that.

---

[1] Note that throughout this paper, double quotation marks ("") are used to indicate *direct quotations* at all times, whereas single quotation marks ('') are used to indicate terms or some bracketing, i.e., a focus on the work—i.e., the commonsense practical reasoning of members of the setting—that is employed to produce the stability of various objects of interest— phenomena—that are under inspection in this paper [34, pp. 30-33]. For example, see 'solving this trouble'.

Adjunctly, this paper's topics also resonate with broad extant concerns in HCI that the craft of professional design practitioners in industry[2] remains under-discussed [99]. There is, fortunately, a growing focus in HCI on investigating this craft (e.g., [38, 41, 42, 43, 40, 66], but also see [8, 97] for precedents). The study in this paper seeks to contribute to this strand of work too, which, I would argue, aims to mature HCI's academic discourses around design practice by shedding some light on what practitioners of various forms *actually* do.

The rest of this paper develops these ideas as follows. The next section offers a brief summary of the large body of research literature that discusses usability testing (and by extension, usability evaluation), examining the methodological preoccupations of this work, usability testing's conceptualisation in HCI, and reviews instances of empirical studies of usability testing as a practice. The paper then introduces and presents some data fragments drawn from a broader video-based study of usability testing at several organisations. These fragments let us take a look at how usability findings are arrived at or, somewhere along the way, 'dissipate'. Finally the paper concludes with a discussion that summarises the key findings, and offers reflections on several overlooked features of usability testing that are seemingly absent from the current purview of the literature.

## 2  LAB-BASED USABILITY TESTING IN HCI RESEARCH AND UX PRACTICE

The gamut of usability evaluation methods have been subject to extensive research in HCI and connected fields, from formal examinations of analytical model-based methods like the Keystroke Level Model (KLM) [12], inspection-based methods such as cognitive walkthrough, through to empirical methods like usability testing. The notion of usability itself has also been a rich topic for HCI but remains subject to considerable conceptual challenges [93]. By focussing on *practitioner-led* lab-based usability testing as an instance of a usability evaluation methods in practice, I seek to set aside these various theoretical discussions of usability that HCI has been preoccupied with and instead consider practitioners' work and their orientations to usability instead [77]. These are likely to be substantively different to usability testing performed in the course of academic research. Nevertheless, a summary of some HCI research is still useful to the conceptual claims made in the course of this paper.

Lab usability testing forms a significant and complex body of research in and of itself, as well as being the subject of various practitioner texts. There is often little distinction made between usability testing as practiced in academic environments compared with how it takes place in industry settings. Further, lab-based usability testing in industry is often described as 'user testing', which, curiously, seems something of a misnomer, given that evaluators frequently stress that it is the artefact that is under test rather than the participant. This leads to a flexibility of the term's application: 'user testing' may be used to refer to a wide range of variations beyond lab-based testing of the sort under investigation in this paper. This includes remote user testing, in situ testing, Rapid Iterative Testing and Evaluation (RITE [68]), and guerrilla testing [3].

Before unpacking some apparent absences in current HCI research on usability testing, I want to first pick out key topics the literature furnishes us with. To begin, much discussion revolves around aspects of usability test procedure. This ranges between the introduction of new ways of conducting such procedures through to critical discussions of their reliability. For instance, the number of participants involved in a test and its relationship to the production of usability findings has been the subject of much debate [72, 88]. It is also a regular topic in practitioner discourse [87, 64]. Equally important and perhaps more extensive has been the topic of 'think-aloud' protocols and their implementation in testing [53, 10, 73, 6, 15, 52], along with some interest in the divergences between academic and practitioner uses of such techniques [67]. Boren and Ramey [6] in particular offer one of the most detailed accounts of what testing involves from the perspective of moderator-participant interactions. The reliability of testing and those performing testing is a frequent concern [64, 69, 70, 54] as is the broader rigour, consistency and uniformity with which methods are implemented [6, 44, 102, 18, 93]. Comparability of usability test results across different evaluation methods has come under scrutiny generally, and usability testing has not been exempt from this [55, 102, 44]. Critical discussion has located differences in the findings of analytic methods (e.g., Goals, Operators, Methods, and Selection rules (GOMS) [13]) or so-called "discount" inspection methods (e.g., cognitive walkthrough) as compared with problems identified during actual usability tests [60, 18]. Much of this debate turns on conceptualising analytic and experimental methods in terms of hypothesis testing [64]. Concerns also extend to how findings are organised and sorted [51] as well as how the move from usability testing to the presentation of findings is achieved by practitioners [26].

---

[2] Some clarity of language needs to be exercised here. HCI research often is unspecific in its reference to aspects of this discourse, meaning that binaries such as theory vs. practice and research vs. practice are often left to blur into one another (e.g., in calls for bridging these binaries; see Beck and Ekbia [2] for a review of these debates). Many times it is left unclear whether HCI literature is referring to 'practice' as synonym for 'industry', or 'practice' to mean 'practic*es*' (in the sense that we *all* have practices of some kind or another), or to mean 'practice' as a broad opposite to just 'theorising', or 'practice' as in 'having a design practice' (which itself remains unclear: academics doing design research often have their own 'practice' even though the outcomes of this practice remains within the confines of academic communities).

Much of this literature offers contributions that are aimed at improvements or modifications to usability testing approaches. For example, this includes different protocols for test moderation (e.g., "question-suggestion" as opposed to think-aloud [45]), reproducing pre-identified crowdsourced problems as part of tests ("creative sprints") [37], ways of compressing testing and the reporting of findings into a single-day model [56], formally integrating design responses in the course of testing [68, 64], improvements to usability problem description forms [60], or developing new training methods to prepare novice usability testers (e.g., to incorporate software developers into testing processes [9]). Significant programmes of work such as the "Usability in Danish Industry" (BIDI) project have sought to establish novel forms of usability testing via the integration of Scandinavian participatory approaches [7, 32, 27, 28].

## 2.1 Conceptualising usability testing

There is a tendency within *some* of the literature summarised above to frame usability testing in particular as vulnerable, and to be in danger of becoming a 'compromised science experiment' particularly for practitioner-led applications of testing. This is reflective of currents of HCI thought. Research focussing on design processes (e.g., user-centred ones) can sometimes tend to view "method as prescription" and construct oppositions between "'proper' use in the academic literature" and "*in situ* use in practice" [40]. For instance, Boren and Ramey argue for the theoretical grounding of usability testing practices, pointing to deficiencies in the conduct of usability practitioners in moderating tests [6]. Nørgaard and Hornbæk, who do not seek to "reprehend the practice of usability testing", nevertheless still sound a note of concern over practitioners' work regarding absences of systematic analysis and potential bias[3] (e.g., in the anticipation of usability problems and their shaping the design of test tasks) [73]. In some ways it is unsurprising that this framing is applied to industrial applications of usability testing, since a significant portion of HCI research *also* treats usability testing performed in academic research environments in a normative scientific frame: one need only refer to the well-known "damaged merchandise" discussions of the 1990s [44, 74] in which this orientation to 'usability as normal science' has been litigated (and the confusions around "validat[ing] discount methods" [16]).

---

[3] An ethnographic note grounded in my studies of practitioner work is instructive here. I have noticed that practitioners themselves *are* often keenly alert and sensitive to this kind of criticism, and often explicitly account for it. For instance, in explaining my interest in studying practitioner work practice, or in video capturing usability testing, practitioners often offered accounts of their practice as 'cutting corners' or 'not doing it properly', and that in examining their work I would probably find it to be lacking in rigour (of course in suggesting such things they were also displaying an orientation to my status as an academic).

That said, conceptualisations of usability evaluation are developing in HCI, and it would be unfair to imply a uniform view. These shifts have pushed towards a more nuanced view of design methods. Hornbæk, for instance, questions various "dogmas" associated with evaluation methods; several of them connect with the topics of this paper [51]. Firstly, he identifies a tricky relationship between a method and the assumption that it "directly leads to identification of problems" (thus also questioning the role of method itself as "guarantor-of-design" [40]). Studying findings' production work in usability testing can help give clarity to this. Secondly, Hornbæk identifies a dogma related to the separation of evaluation and design: although usability testing and design process in practice go hand-in-hand, strangely, usability testing sites do not routinely seem to be taken equally as sites for design (although there *are* approaches that formally integrate design into testing in some manner [68, 92, 27, 28]). Hornbæk suggests instead that "usability evaluation is really idea generation" (also see [50]); this paper offers empirical evidence for this notion, albeit painting a more complex picture still. The point is that notwithstanding their correctness, such questions around fundamental "dogmas" only increase the need for HCI research to take a sustained and close look at how findings from usability evaluation methods come to be practically produced, as an equally important component in—and inextricably entangled with—formalised accounts of such methods in the literature.

## 2.2 Empirical studies of usability testing

Given the extensive literature on usability testing, what of empirical studies of testing as-it-happens, as a collaborative work practice? Although some HCI research does focus on observations of practitioners performing usability testing (e.g., [73, 86]), it eschews interactional detail of the kind presented in this study. In other words, there is no close examination of usability testing as a socially organised phenomenon, i.e., as a concerted, sustained, moment-by-moment, unfolding *interactional* achievement of the various stakeholders engaged in testing[4]. And although there are practitioner accounts of usability testing, such as Rosenbaum's historical tour of the evolution of usability testing in industry settings [78 pp. 344-378], these tend to be broad brush and do not present detailed 'naturalistic' inspections of usability testing activities. One possible exception is a paper by Bowers and Pycock that examines problem formulation [8]. However, although this work presents an account of moderator-participant interactions, the situation studied is that of cooperative design (with a participant who seems to be a ratified member of the design team), rather than a more typical usability test involving externally recruited participants, stakeholder organisations, and a traditional lab configuration.

Beyond HCI, there are existing sociologies of testing [75] such as Woolgar's well-known paper on "configuring the user" which examines usability testing and its surrounding procedures [101]. However, even then, such studies tend to avoid any serious *praxeological* focus on testing itself—that is, on the methodic practical actions of members of the setting. While Woolgar *does* articulate a constructive sense of 'being a participant'—that the user participates in the achievement of the test and its outcomes—his study steps away from unpacking members' analyses of actions that make the test what it is. Instead it seeks to cast usability testing within the frame of contemporary sociological theorising [11] by discussing the artefact being tested as a "text" that is "read", thus locating it within "longstanding problems in social theory about agency and object" [101]. This then does *not* provide us with much purchase on how participants and stakeholders in the test analyse the test itself and produce findings.

---

[4] Note that throughout this paper I am using the term 'stakeholders' to include UX practitioners, clients, designers, developers, etc., but *not* to refer to participants. This is a matter of terminological clarity for this present paper, although it is worth acknowledging that some practitioners may include participants when they refer to 'stakeholders'. Use of this term also underscores that testing can be thought of practically as "stakeholder-centered design" [25].

Curiously, awareness of and complaints about the absence of detailed studies of usability testing as a work practice are relatively easy to locate in HCI research. This includes remarks on the deficiencies of instructional works about performing usability testing, as well as deficiencies of present (HCI) research on the subject. Drawing implications from their study of practitioners' work, Nørgaard and Hornbæk point out that the "practical realities influencing tests are much more frequent and severe than one would expect from textbooks or research papers on usability evaluation" [73], something Lallemand and Koenig concur with [59]. Følstad, Law and Hornbæk elaborate on this, noting that "Texts on usability testing [...] lack detail on how to do analysis" [29]. Now, while *some* textbooks are light on the detail of what happens during a test (e.g., see [94, 58]), others actually do provide quite detailed guidelines for observation (e.g., [23 ch. 6, 24 p. 293, 1 ch. 7]), often offering step-by-step 'script' or tutorial approaches to assist novice moderators through elements of usability testing, exemplified by works such as Krug's popular guide [57].

We can choose to read such concerns in different ways. The first way is as confoundment over the divergence between formal accounts of testing and the actual observed practicalities of testing, i.e., complaints that textbooks and papers documenting methods do not match up with the realisation of those methods in practice, and, furthermore, that *this* is a problem (a concern also noted in related design research that contrasts "modeled understanding of design activity" with "authentic design practice" [40]). However, we can dissolve this problem quite easily by better understanding the relationship between formal accounts and the practical circumstances they relate to. In this alternate view, such discrepancies are instead an *irremediable* feature of methods descriptions (as rules, instructions, procedures, recipes, etc.). Formal accounts of usability testing methods cannot be ultimately remedied because such accounts always remain incomplete. The reason is that methods descriptions and the actual sense of them—not to mention their in situ adequacy as prescriptions—must be discovered and worked out for all practical purposes *in the course of* their actual accomplishment (which itself also affords rules, instructions, etc. a vital role in the moral order). Thus methods and their use are inextricably connected; formal accounts of methods cannot readily be understood as a separate self-sufficient entity. This is a well-known point located in a thread of thought spanning Wittgenstein on rules and rule-following [100], through to ethnomethodology on instructions and instructed action [34 ch. 6], and, ultimately informing Suchman's well-known distinction between plans and situated actions in HCI [90 pp. 101-104] (see also see [46]).

Having dispensed with this issue, we can instead choose a second reading: that these are concerns about *limitations* in HCI's understanding of usability testing. Kjeldskov, Skov and Stage point out that data analysis in usability testing is only ever "vaguely described" [56], while Nørgaard and Hornbæk state that studies of usability evaluation "rarely describe the process of evaluation in detail" [73]. What is being pointed to is absences of knowledge about the praxeology of usability testing (and not less useful critiques of the efficacy of 'the textbook'). Without a fuller picture, the possible effectiveness and impact of advancements in usability evaluation may be reduced. Følstad, Law and Hornbæk thus call for remedial action that examines "research on evaluation in context" [27]. By doing this, our understanding of what is going on within usability evaluation techniques and their application in industry settings will be improved, but also we will have better conceptual grounds for HCI's attempts to support and improve them. This paper attempts to do just that: to broaden HCI's understanding of usability testing as an evaluation method by precisely spelling out what testing's praxeology actually is for practitioners without embedding a rebuke of the supposed deficiencies of formal accounts.

## 3   CAPTURING PRACTITIONERS' USABILITY LAB WORK

At the core of usability lab work are findings. This can be meant in a quite broad sense to encapsulate the wide range of outcomes that are achieved in the course of testing. Traditionally, locating findings means identifying "defects" [50 p. 267]. However, other matters may be rolled into what constitutes a finding, such as desired features (perhaps labelled as 'insights'). To get at this phenomenon I conducted an ethnomethodological study of usability labs. In the present paper I employ video recorded fragments of these testing sessions to serve as exhibits of the findings production phenomenon. These were captured from the point of view of the observation room that is in some sense 'spectating' on testing as it takes place. While the video data shown in this paper is restricted to only one form of usability testing and one particular working arrangement (the observation of testing certainly does not always involve clients or designers), the study provides a window into how the work of testing is accomplished and establishes a broader agenda along which future studies can be directed. Next I describe how I came to build case studies of usability testing practice, offer detail on the ethnomethodological approach, and, with reference to ethnomethodological considerations, explain how fragments were selected.

### 3.1   Locating study sites

To begin with I must briefly situate the paper's focus on usability testing within the context of a wider investigation conducted as part of a Fellowship funded by the UK's Engineering and Physical Sciences Research Council (EP/K025848/1). Primarily this explored the constitution of industry UX, design practitioners' work, and the connections between this and academic HCI research that ostensibly concerns itself with very similar topics. While establishing a better understanding of industry practice, I initially sought to take in a wide picture of the UX milieu, including activities such as attending numerous practitioner conferences, both international and national (UK), and more informal tech 'meetups', all categories of which I have also delivered research dissemination talks acting as a participant. These activities led to the realisation that a more focussed look at practitioners' work as a material practice was needed, underlined by absences I have outlined in the previous section. Usability evaluation methods in particular seemed fruitful and practicable as an anchor point for that work.

Oriented by this, I adopted a process of attending, observing and video-recording (where possible) usability test sessions in different organisations. In some cases this capture activity coincided with longer-term interactions with practitioners of varied roles over a number of years, discussing their work, project organisation, methods, attitudes, orientations, and so on, all of which formed the backdrop to more focussed studies of practice. Specifically, I performed participant-observation of six instances of usability testing. These were conducted in various UK-based labs: two in UX agencies, which we will call A1 and A2; and in two in-house UX teams of contrasting size, one sited within a large media organisation, IH1, the other embedded within a higher education institution, IH2. During the research I was present for 11 days of testing across the six projects. The form of testing employed ranged between the evaluation of digital and paper prototypes in service of future design iterations (A1, A2), to product baseline testing (IH1, A1), to user acceptance testing (IH2). The artefacts themselves spanned multiplatform media players (IH1), websites (A2, IH2), mobile apps (A1), and web apps (A1). From these sites I collected around 37+ hours of video recordings (i.e., a subset of the actual observation time since access proved to be restrictive). There are various limitations to this observation and capture work that are important to bear in mind: it involved only selected styles of testing (e.g., no remote testing), did not investigate more than one iteration of a given artefact under test, nor did it necessarily span the broadest range of possible types of design projects.

**Details for site A2.** Of this broader corpus, this paper presents fragments drawn from the 8 hours of video data that was captured from site A2 (reasons for which will become apparent shortly). A2 is a small ~6-8 person consultancy located in the UK that focusses on "human-centred design" (their description). The consultancy offers a range of research, design and evaluation services including usability testing. A2 was conducting a two-day usability test for a client organisation; their client was a regulatory body that handles media complaints with its various audiences including members of the public, but also industry such as advertisers, or standards enforcement bodies. The artefact under test was a mid-fidelity pre-launch prototype of a new website for the client organisation. This prototype had been developed by a design agency contracted by the client (delivered during the test via a prototyping tool, InVision); some pages of the prototype were more high-fidelity while others were less so. The stated aims of the test were focussed on a range of elements of the prototype, including submission and tracking of the organisation's complaints handling process, the prototype's overall appearance, and its provision of information about the organisation, including past rulings on complaints, and so on. In this sense, the test was similar to relatively 'textbook' forms of usability testing (e.g., "completing a transaction" and "evaluating navigation and / or information architecture" [94 ch. 3]). Test moderation with participants followed a relatively traditional, formal think-aloud approach. The single-sitedness of the data should be borne in mind, along with the simple fact that A2 offers us only one type of design project configuration (i.e., technology in question, prototype fidelity, etc.). However, more comment on this matter is offered next.

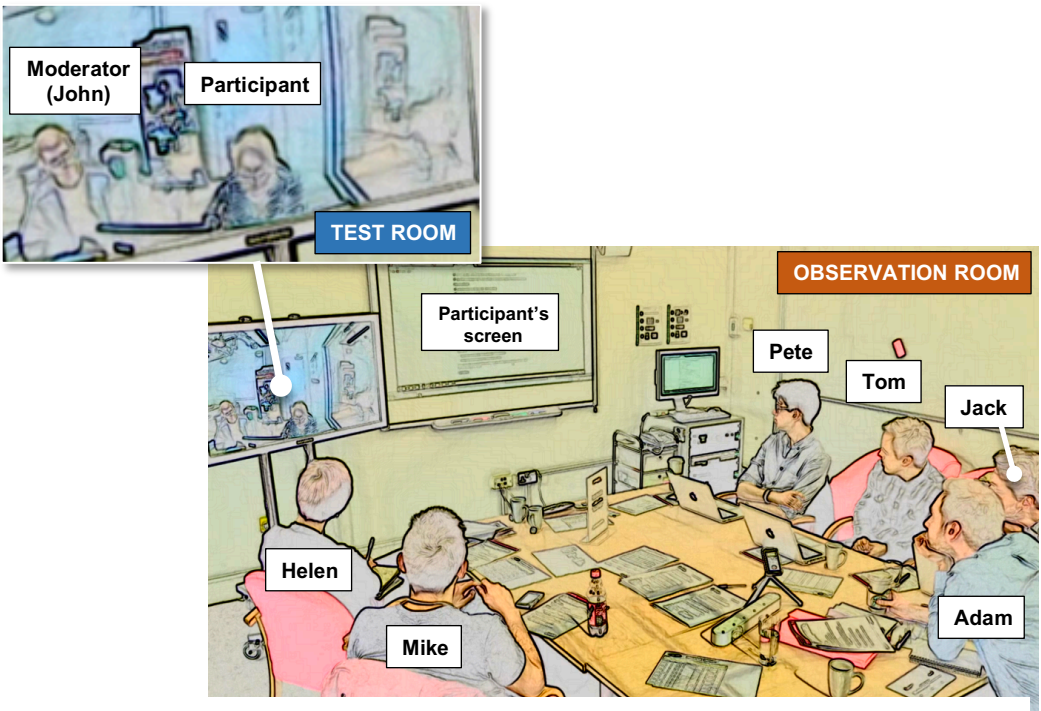

**Figure 1: Site A2. A live view of the test room is visible on a large display in the observation room as indicated with the enlarged cut-out.**

Figure 1 shows the two interconnected rooms at site A2 that I will refer to in the rest of the paper: the test room, where the moderator (John, a consultant with the lab) and participant sit;

and the observation room, where the representative of the client[5] organisation (Adam) observes alongside the three contracted designers who developed the prototype (Pete, Tom and Jack), and the rest of the lab consultancy team (Helen, a more junior consultant, and Mike, a more senior one). Although the designers here do use the term 'UX' in their roles, they have *not* been involved in any usability testing prior to this. 'UX design' is often offered as *distinct* service to 'UX research' (although practitioners may provide both), which is likely the case here. Adam, however, *is* more familiar with usability testing and its procedures, which led to him to commission it. As a client representative Adam acts as a 'champion' of the process within his organisation and subsequently re-presented outcomes of the testing (as documented in the formal usability testing report) to the client organisation in order to support the process of the website redesign. A live video stream is also visible on a large display in the observation room, while the participant's display is projected onto a screen at the head of the table in the observation room. Talk between the participant and moderator is audible via a speaker system. Consultants Helen and John alternated moderating the test for the 14 participants who attended the sessions. Finally, it is important to note that the designers (Jack, Pete, Tom) and client representative (Adam) were present throughout testing, however this is not uniformly the case in industry practice and sometimes other stakeholders may not be present at all (and perhaps be provided in those cases with a 'highlights reel' of the test).

### 3.2 Approach: Ethnomethodology, perspicuity, and generalisation

The study has been pursued primarily from an ethnomethodological (EM) perspective [33, 34, 36] but is also informed by conversation analysis (CA) [79, 80]; often this is amalgamated as EMCA. There is no space to offer a detailed explication of EMCA-oriented approaches, which have their origins primarily in the work of Garfinkel and Sacks (see [36]) and built upon various philosophical and sociological precedents set by Wittgenstein, Schütz, and others [96, 85]. Nevertheless, although still a minority perspective in HCI, EMCA does have significant historical currency in CSCW and HCI, most notably with Suchman's critical appraisal of the role of cognitive science in HCI design [90]. EMCA approaches have also provided fruitful ways of examining workplaces as sites of collaborative action with and around technology (e.g., see [47]). In what follows I refer specifically to EM positions, however the reader should note that many of these are nevertheless shared with CA, although CA has historically concentrated mostly on *talk*-as-action.

EM concerns itself with explicating the ways in which members of society, as practical 'everyday sociologists', continuously attempt to create and maintain a stable and orderly social world, and do so in a way that is designedly 'accountable' to the situations they find themselves in. EM's preoccupation is with articulating the *methods* that such members employ to bring about that stability and orderliness, i.e., what Garfinkel calls "members' methods" [33 p. vii]. In this case, our members are participants, user researchers, designers, and any other stakeholders engaging in the work of the usability test. Because of EM's interest in methods and their natural accountability, the reader will see in what follows various detailed descriptions of them as they come to be deployed by members of the test so as to bring about *its* order and as a matter of this to make that order jointly recognisable as 'doing usability testing'. To elaborate this particular ethnomethodological view, we might say that social actions, as methodic and accountable actions are designed by members with an "essential reflexivity" [33 p. 7][6]. As Sacks puts it, this means that "actions [are...] done in such a way as to be recognized, so that the apparatus that can provide for how they're recognizable can also constitute procedures for generating the occurrence as a recognizable action, or set of them" [79 vol. 1 p. 237]. There is an important 'circularity' to this idea in that actions—such as noticing and topicalising some participant activity as a usability problem—must contain within them adequate accounts of what kind of action they are. This is an idea which will be explored throughout the paper when considering how members of usability test settings work to surface and produce findings and in so doing simultaneously provide an account of their provisionality and relevance.

---

[5] There are, of course, other possible meanings to 'client' and a range of complexities for agency-client relationships that this paper cannot cover but must nevertheless be acknowledged (for practitioner accounts, see [71]).
[6] Caution is required here because EM's notion of reflexivity differs substantially from other uses. Lynch describes it in the following way: "The reflexivity of accounts [...] alludes to the embodied practices through which persons singly and together, retrospectively and prospectively, produce account-able states of affairs." [63] Ethnomethodological reflexivity

EM-informed work has little interest in engaging in theoretically-driven descriptions of social actions, but instead seeks to concretely articulate the ways in which members of a given situation *actually treat* one another's activities in the course of concerted joint action (in other words, categorising activities as this-or-that kind of object, and, as per reflexivity, building in a stance on that categorisation in ones' own next actions). In taking an ethnomethodological approach to investigating usability testing labs, I am primarily interested in locating, excavating and then reporting the various ways in which members of the setting accomplish testing work—the locating of findings—and *in so doing* (reflexively) show that accomplishment for what it is, i.e., how social interactions come to constitute findings as recognisably adequate objects. In this sense EM has much to offer studies of usability evaluation given that this aspect—the "seen but unnoticed" [33 p. 36] practical achievements involved in doing testing—tends to be absent from the HCI literature.

Some things must be said on the perspicuity of usability labs as well as arguments regarding generalisation in EM given the focus here on site A2 only. Garfinkel's notion of the "perspicuous setting" [34 pp. 181-182] is relevant to better understand how we might *see* the work of usability evaluation in action, as a praxeological endeavour. Usability lab spaces offer us perspicuity in that labs' raison d'être *is* to enhance the visibility of participants' problems when interacting with or otherwise responding to artefacts under test, and then make those problems readily available to observers. Now, local practices of different observation rooms in usability testing environments do vary, and A2 was selected for clarity, yet is also not intended to be generically 'representative' (in line with EM or CA arguments, see [21] and [82] respectively). For instance, some tests may involve mostly (IH1) or entirely (IH2) silent observation work. In the case of IH1, for example, sticky notes were primarily employed during test sessions to produce initial findings, which were then subject to a process of affinity diagramming or 'triaging' both in between sessions and at the end of testing entirely, all in order to produce a set of agreed-upon findings to be formalised in the reporting phase. Thus, instead of claiming the validity of A2 as a generalised 'typical' or 'average' sample, A2 simply offers us—in its happenstance—better access to and clarity of exhibits of usability findings' formulation work (hence, acting as a perspicuous setting). Of course, other sites necessarily *also* turn on such formulation work, although this work is typically being done somewhere else (although it is worth noting that at site A2 there were also more 'private' activities taking place alongside these such as note-taking on laptops).

Finally, specific site selection also presents certain limitations in that description of the diversity of approaches to usability testing remains pending. Nevertheless, there is a strong practical advantage that recommends offering just one site to the reader, in that it (hopefully) helps retain coherence as well as an improved sense of readability.

### 3.3 Approach: Fragment selection

---

also has "radical" tendencies as much as "endogenous" ones [76] in that ultimately EM "recognizes that the accounts of order it produces, are, themselves, reflexively embedded within the context in which they are produced" [96 pp. 117-8]. Whether this is significant or not is a matter of debate [63] dependent upon what 'doing being reflexive' actually leads to in terms of intellectual outcomes (e.g., see Livingston's analysis of CA data sessions [62 ch. 6]).

How were fragments selected from A2's video data? To begin with, ethnomethodology argues against any prescriptive methods being applied to such investigations as this runs the risk of potentially constructing entanglements and confusions between members' methods (i.e., that EM is soley interested in) and researchers' own analytic methods as located in various extant social scientific-theoretic frameworks (e.g., categorisation schema, topical focus, etc., see [11]; also see Cicourel's critique of social scientific methods [14]). This is not to say ethnomethodological investigations are somehow in competition with other approaches in the human sciences but rather that its investigations operate on a different—or, as vom Lehn points out, "alternate" rather than "alternative" [96 pp. 25-26]—basis.

An important factor in fragment selection is how the wider investigation informed the approach to A2 as a site. To begin with, selection from A2 was broadly based in the observation of a grossly noticeable—not to mention fundamental—feature of usability testing present across all sites. This was as follows. While *certain* features of the test were treated as 'problematic', 'troublesome' or simply 'of interest' by those observing, *other features*—which might in another time and place reasonably have been something of concern to testers—were ignored or dismissed in some way. Usability lab discussions between stakeholders—whether conducted during the test or mostly afterwards as part of impromptu meetings—were organised in such a way so as to assess test proceedings, weighing them against a range of contingent matters, and out of all of this, surface particular features as relevant and thus to be taken forward in some way. It was through participating in the usability testing sessions myself as something of a naïve observer that I was able to appreciate firsthand this matter of both seeing and then sorting the relevant from the non-relevant as a key challenge and accomplishment for stakeholders, something that was made all the more pertinent when stakeholders sought to engage me with questions about 'what I thought' or other ways of eliciting my input. This is notably *different* from a 'detection of invariant properties' conception of usability evaluation methods mentioned earlier. In this view, problem identification rises asymptotically with the number of test participants or expert evaluators (hence single case usability problems being a challenge [30]). Testing's work is ultimately a matter of sorting the 'signal' from the 'noise' to reduce Type I and II errors, i.e., false positives / false negatives [64]. However, I argue that the accomplished nature of what even constitutes 'trouble' calls this position into question.

Operationalising this basic interest for A2's video data meant performing an initial 'logging' phase (also carried out for IH1, IH2 and A1). This meant that all captured video from A2 was examined, with activities documented in order to re-orient myself to what happened during testing and capture what was available on the record[7]. Guided by the framing outline above, I sought to collect together different moments during A2's testing where stakeholders in the observation room located something of relevance happening in the test room and *responded* in some observable way. Within this broader set of moments, I then concentrated on unpacking in more detail those interactions where stakeholders presented assessments of what was happening in more extended ways (as this broke open for availability and presentation the kinds of work entailed in findings' production). This in turn produced around 26 fragments that were transcribed in as much detail as those presented in this paper, and from which the present set was selected primarily for clarity to the reader. In the course of this work, multiple "data sessions" [48 p. 102] with colleagues were also held so as to examine (in detail) particular video fragments with other EMCA oriented researchers. (Note that the fragments here use an adaptation of Jeffersonian transcription [49]; in short, this transcription format preserves features such as overlaps, pauses, as well as matters of prosody, amongst other things.)

---

[7] Of course, video recordings themselves do analytic work: they render certain visibilities but omit other features; also see [48] for more detail.

## 4 THE WORK OF PRODUCING USABILITY FINDINGS

Fragments in the following sections show stakeholders practically 'sorting through' and categorising matters they observe occuring in the test room as this-or-that kind of object. The arrangement of fragments A-D is designed to help present, in a staged way, the outlines of that complex interactional work. Collectively the fragments highlight only the main interactional 'trajectories' through this (so does not claim to be comprehensive).

As I have alluded to, the fragments also show how the work of producing usability findings turns on the ways in which members of the observation room treat phenomena occurring in the test room as *troubles* in need of *solutions*. My use of this language is quite specific for the following reasons. Firstly, when we talk about the stakeholders 'looking for' and 'seeing' trouble, what constitutes that trouble is jointly and unfoldingly shaped up *by them*; as per the position detailed thus far, 'trouble' is *not intrinsic* to some aspect of an artefact under test, a participant's conduct, the skill of the moderator, and so on. This means that trouble in the observation room—the OR—(as we will see) is *not necessarily* the same as any articulation of trouble that might emerge within the test room—the TR—between a moderator and participant. Secondly, by using these notions of trouble and solution I am also stepping back from the term 'findings', as commonly referred to by practitioners. 'Findings' is fraught, since a finding a) may begin life *initially* as a mere implication standing silently alongside what are explicitly stated solutions in-test, and b) findings might subsequently come to be produced as identifiable things only in a post-hoc 'reverse engineering' process performed for the purposes of formal reporting. Thus, employing trouble and solution allows us to be more precise about what is being done by practitioners, and at the same time preserve and therefore also inspect the sequential *ordering* of troubles and solutions (see Fragment D: Discussion).

### 4.1 Looking for trouble

If the main activity for stakeholders in observing usability testing is formulating findings, then the corresponding problem to be continuously resolved by them is that of the relevancy of participant troubles. By relevancy I mean (somewhat simplistically at this stage) determinations of whether TR proceedings are indeed offering something that prospectively *might* be shaped into a formalised finding within some form of usability report. This matter of the relevance or non-relevance of what is going on in the TR is thus strongly underlined by the often tentative, vague, implicatory, elliptic, euphemistic, provisional, and deeply indexical ways in which stakeholders topicalise what just happened or is currently happening in the TR. In other words they work to bring TR events into interactional, often conversational (topical) relevance or focus for co-present others. This is a practice I am calling 'surfacing'. By surfacing TR events in some way, stakeholders can enable TR proceedings to get 'seen'—to be made visible—in the OR as trouble of this or that sort. As part of surfacing trouble, there are a range of contingent matters at play that feed into stakeholder assessments of relevancy, which I refer to as 'relevancy devices'.

Stakeholders co-orient to a great many different kinds of seen, witnessable trouble that unfold in the TR and ultimately feed into the formulation of findings in one way or another (typically via a formal report, but this may also be in the form of a slide deck). Nevertheless, for much of the time, both in the setting I present here but also for sites A1 and IH1, stakeholders mostly tended to sit reasonably silently while watching and listening to the TR's live video feed until something striking them as relevantly 'troublesome' emerged, thus warranting an initial surfacing action of some form or another (such as a glance, or Jack's "interesting", below) that then usually occasioned some discussion. Note that Fragments A-D only concern what one might call defects and *not* other types of findings that can emerge from testing (such as participants offering generative accounts like suggesting a new feature or praising elements of the test artefact in some way).

Although it is *not* the focus of this paper to offer an exhaustive list of different types of trouble that subsequently come to be articulated as usability findings in this way, I want to offer a broader sense of them to frame what comes next. Hence:
1) participants might be seen as get 'stuck' in some way during a task, i.e., unable to proceed (typically paired with an account of the trouble they are encountering);
2) participants could be seen to 'go off task', i.e., engaging in activities that diverge from and are in some way orthogonal to a normative sense of 'doing the task';
3) participants might—as if 'independently' of their current task—produce an account of a perceived deficiency, a preference or opinion on the artefact;
4) a participant may be seen to be inadvertently revealing a deficiency of some sort that is neither 'getting stuck' nor 'going off task', e.g., a trouble noticed in the OR but does not *practically* impede the participant completing the task or necessarily occasion an account of trouble on their part (an example might be inconsistent design).

In the next two sections I present two differing fragments. The first shows a fairly straightforward moment where the OR *formulates* a finding and which subsequently gets articulated in the formal usability report issued by the UX agency after the test's completion (this formal report divided each section into "insights", "issues" and "recommendations"). The second fragment shows the inverse: a moment where some kind of trouble is raised in the TR, but while the OR does take this up, it is done in such a way as to make it non-relevant and is thus *dissipated*.

### 4.2 Formulating a finding

As this paper concerns practitioners' work, a critical element is in the role of organisational concerns—particularly the constraints and requirements of a client—during formulation work. Usability testing, as it is practiced in industry, is necessarily riven with such things as potential matters of relevance that may be brought to bear at any given moment on findings' formulation. Since a very wide range of these organisational concerns may surface during testing work, we can only scratch the surface in this present paper. In providing an exhibit of the emergence of trouble (of type (3) above in the previous section), this fragment naturally highlights only *one* of the many ways in which such wider organisational concerns faced by industry practitioners may be deeply woven into the moment-by-moment unfolding of production work (here, human resources within the client organisation). This fragment also offers an initial sketch of usability testing work practice, and thus acts as a familiarisation tool for understanding the progressive, collaborative way in which members of the OR see and then treat trouble.

**Figure 2:** "Make a complaint" form, login or register section

**Fragment A ("we just haven't got the resource"): Description of transcript.** As we join the action in the TR, Helen (M) is moderating the session with Participant 6 (P6). P6 has arrived at the 'login' section of the complaints form (Figure 2; see Figure 6 for the beginning of this form). Filling in the "Make a complaint" form is one of the central tasks of the usability test, and the form has various stages that each participant is presented with (Fragment D shows the start of this particular task). The part of the form featured in this fragment is reached once the user gets past earlier stages of filling in various details about the nature of their complaint (including where they saw an offending advertisement and when, what medium, and a description of the complaint). The transcript below begins just as P6 is reaching a 'continue' button that then reveals the next stage of the form (as seen in Figure 2). The following is divided into parallel transcripts, with actions in the TR and OR presented in a way that preserves the simultaneity and timing of activities in both rooms. The transcript is first presented below in full and subsequently broken down in to pieces for the following analysis to aid comprehension of the organisation of OR activities.

|  | **TEST ROOM (TR)** |  | **OBSERVATION ROOM (OR)** |
|---|---|---|---|
| **P6** | yeah so: it says (you see) right here (1.2) so I'm guessing that's done and I continue | 01 02 03 | |
|  | (0.5) | 04 | |
| **M** | (okay) | 05 | |
|  | (3.5) | 06 | |
| **P6** | no::: (1.3) log in or register (0.7) no don't want that (2.2) hh that's a bit annoying (1.2) ((turns to M)) I hate logging (in) and registering to every (little) (  ) so: (0.3) if I made a complaint I'd rather you just asked for my details and (then) contact me rather than me have to log in beca- as if I'm gonna make an a- complaint every week (0.7) I think logging in and registering makes me feel like (1.3) I'm ah regular u(huhu)ser(uh) (.) and so I'll be complaining about something every five minutes | 07 08 09 10 11 12 13 14 15 16 17 18 19 20 21 22 | 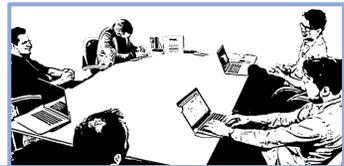<br><br>**Adam** huhh<br>((Adam begins glancing between Mike, Jack and Pete; Mike and Pete glance towards Adam; quiet group laughter))<br><br>((laughter<br>  in<br>  room<br>  and more glancing)) |
| **M** | okay | 23 | |
| **P6** | so I just prefer like something that says (.) email address (.) contact number (0.3) address something like that and then (1.4) hopefully someone will get back to you that way<br><br>**<REMAINING TALK IN TR OMITTED>** | 24 25 26 27 28 29 30 31 32 33 34 35 36 37 38 39 40 41 42 43 44 45 46 47 48 49 50 | **Adam** ((turning to Jack)) we need to get rid of that ((Jack nods))<br>**Mike** ((glances at Adam))<br>**Adam** ((turns to Mike)) because w- we actually want to deter that, we don't want people logging in:: (.) and registe:r: (.) a complaint every week (.) cause<br>**Mike** yeah<br>       (0.8)<br>**Adam** >if people< (0.6) I'thig usually if like t- if people are (limi-) (like) ten or so: like within in a short space of time we'd almost like kindof<br>       (0.7)<br>**Mike** ignore them<br>**Adam** yeah (.) >(       ) but we'll< (.) kind of jus' tell them that we c'n see your complaint (in a rush)<br>**Mike** [yeah<br>**Adam** [in a polite way cause (0.9) we just haven't got th- (.) the resource to basically deal with people almost acting as like (0.7) police themselves and just=<br>**Mike** =yeah |

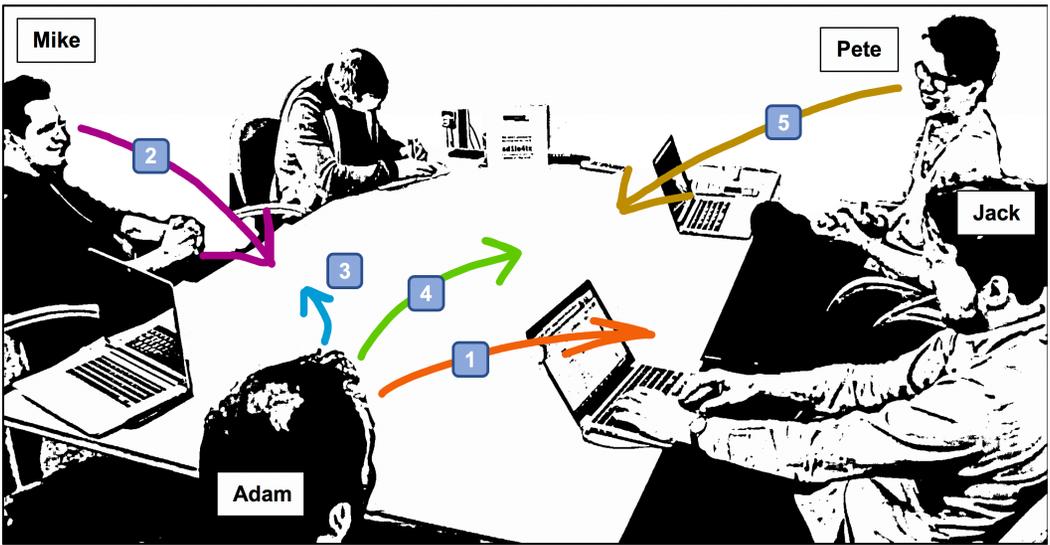

**Figure 3: Glancing between stakeholders (approximate sequential order of glances indicated)**

Trouble in the TR is at first surfaced by several members of the OR in a provisional, elliptic way, emerging as scattered moments of shared laughter and coordinated glancing between stakeholders. Specifically, as P6 remarks "that's a bit annoying" to the moderator, over in the OR, Adam (the representative of the client organisation) begins looking around at Jack (designer) and Mike (UX lab manager), who then reciprocate glances, Figure 3. Meanwhile in the TR, P6 carries on with her account to the moderator. The sequence of glancing depicted in Figure 3 is also embedded in the following transcript excerpt at approximate initiation points:

```
P6    ((turns to M)) I{1}hate {2}logging (in)    10   Adam   huhh
      and registering{3} to every (little)       11          ((glancing and quiet
      ( {4} ) so: (0.3) {5}if I made a           12            group
      complaint I'd rather you just              13            laughter))
      asked for my details and (then)            14
      contact me rather than me have to          15
      log in beca- as if I'm gonna make          16          ((laughter in room and more
      an a- complaint every week                 17            glancing))
```

During this response by P6 to the login / register section of the complaints form, there is sporadic collective laughter in the OR. Framed by the prior glancing between Adam, Jack and Mike, this collaboratively produced laughter suggests a developing, mutually-emerging alignment between members of the OR to the possible relevance of P6's account as a relevant trouble.

Adam's initial utterance ("we need to get rid of that") arrives some time later after this laughter and glancing:

```
P6    so I just prefer like something           24
      that says (.) email address (.)           25   Adam   ((turning to Jack)) we need to get rid
      contact number (0.3) address              26          of that ((Jack nods))
      something like that and then (1.4)        27   Mike   ((glances at Adam))
```

Quite tersely to begin with (as we will see with Fragment D), this utterance by Adam formulates a possible *solution*: the need to remove an unspecified "that" in the design. Interestingly, the incompleteness of this utterance is treated as adequate by other OR members. It is also sequenced with Adam's glance towards Jack, thus configuring Jack's relevance to this utterance and implicating the solution as a matter of prototype design (Jack is, after all, one of the designers). Adam's "we need to get rid of that" indexes an assumption of shared understanding between OR stakeholders: i.e., a commonly oriented-to sense that something about P6's conduct is transparently troublesome. Adam's solution here also leaves the actual trouble initially implicit, perhaps because P6 has already articulated that trouble herself. To reiterate, however: trouble can be formulated in the TR, but is only ever a resource for the OR's own trouble formulation work. Thus it is notable that members of the OR do not entirely adopt all of P6's original account of the trouble, such as her rejection of being categorised as a "regular user".

It is only when Adam turns to Mike (who has already sought Adam's gaze previously), and as the sequence between them then unfolds, that we gradually see Adam's initially terse formulation of a solution turn into a fully developed account of the trouble. Critically for this fragment, Adam's account happens to turn on the working practices of the client organisation.

```
25   Adam   ((turning to Jack)) we need to get rid
26          of that ((Jack nods))
27   Mike   ((glances at Adam))
28   Adam   ((turns to Mike)) because w- we
29          actually want to deter that, we don't
30          want people logging in:: (.) and
31          registe:r: (.) a complaint every week
32          (.) cause
```

At first this begins with a revision of Adam's initial solution formulation, from "get rid of that" to "we actually want to deter that". Mike and Adam maintain their mutual gaze and Adam then continues to expand on "deter" with a comment about not having users register "a complaint every week".

This is subsequently developed into an account by Adam that brings his organisation's practical arrangements to bear, specifically how *frequent* complaints from individuals are handled by them. Adam says "we'll [the client organisation] just tell them [the complainants] [...] we c'n see your complaint".

During this Mike offers agreement and other demonstrations of alignment, such as by completing Adam's truncated utterance with "ignore them". This then leads to Adam introducing 'human resources' ("we just haven't got [...] the resource") as a feature of the organisational process of handling complaints, i.e., that his organisation cannot field enough resources if users of the site choose to act as "police".

```
35   Adam   >if people< (0.6) I'thig usually if
36          like t- if people are (limi-) (like)
37          ten or so: like within in a short space
38          of time we'd almost like kindof
39          (0.7)
40   Mike   ignore them
41   Adam   yeah (.) >(     ) but we'll< (.) kind
42          of jus' tell them that we c'n see your
43          complaint (in a rush)
44   Mike   [yeah
45   Adam   [in a polite way cause (0.9) we just
46          haven't got th- (.) the resource to
47          basically deal with people almost
48          acting as like (0.7) police themselves
49          and just=
50   Mike   =yeah
```

**Fragment A: Discussion.** We can begin by observing that there is a dual aspect to TR-OR organisation that is a frequent feature of usability testing setups. In both rooms there are conversations taking place: between participant and moderator in the TR, and between various stakeholders in the OR. Yet they are connected by an asymmetric pairing where the OR responds to things that the participant says and does in the TR. Being asymmetric, these are not typical conversational 'responses' between members of the two spaces but rather attempts to select and render particular moments witnessable in the TR audiovisual stream to shared relevancy in the OR. On the surface this fragment offers us a fairly straightforward instance of one such response: of 'something' being noticed by members of the OR in their own ongoing analysis, within the frame of this TR-OR relationship. Formulating the trouble emerges in the OR here as a matter of moving on from the initial noticing of the 'something'. Remember, these are troubles as 'seen' by members of the OR. They are matters unfolding in the TR that OR stakeholders are orienting to *as troublesome* and thus are deemed to warrant some kind of account of them (whether that be a glance or a design solution).

The trouble seen here by the OR—in their accounts based around the login and register elements of the form being problematic for participants—is progressively 'worked up' both in terms of an explicit articulation of just what that trouble is, and a discussion around possible ways of dealing with that trouble, i.e., candidate solutions. In looking for trouble, the OR can indeed take the articulation of trouble by a participant and transform it to a *different* kind of relevancy. In this case, the participant presents a personal dispreference by presenting resistance to affiliation with a particular membership category, namely not wanting to be seen as a regular user of the site. In the context of media complaints, this resistance is employed to avoid the implicative 'regular user' = 'regular complainer'. P6's laughter-inflected "u(huhu)ser(uh)" also accounts for that resistance by treating the proposition of being seen as a regular user as comical. Instead the OR reformulates this *personal* categorisation P6 is seeking to contrast herself against to an *organisational* issue about managing the volume of complaints (which then itself leverages a different membership category, casting some potential users negatively, as "police"). The nature of this transformation of TR-generated trouble to the OR's rendering of that describes something of a loosely-coupled connection between TR-trouble and OR-trouble that is structured by a given team's social organisation.

In this fragment we also see how *the practical workings* of the client organisation may themselves become built into the OR's progressive realisation of that trouble and its solution. Drawing attention to Adam's use of "we" across this fragment lets us see how this is achieved [79 vol. 1 pp. 148-149]. Adam's initial "we" (line 25) is sequenced carefully within his glancing at Jack and reciprocation of Mike's gaze. Alongside this, Adam's shifting sense of who "we" indexes as a categorical affiliation enables him to offer, at various points, different ways to hear the sense of that "we" in his developing accounts. Hence, at times "we" may be used to index members of *this* usability test, or "we" as members of *this* project, whereas at others (and most importantly for this fragment), the sense is "we" *as* the media complaints handling organisation and the various personnel, processes, mechanisms, protocols, policies, and so on, that such an organisation's work might entail. These kinds of moves by Adam enable him to manoeuvre from the preliminary surfacing of the trouble that "we"—the assembled members of the OR—have just seen, to a "we" that begins implicating the internal guts of the organisation as critically relevant to the assessment of this particular trouble. This use of "we" then offers the structural support for Adam's unfolding account of current processes of the client organisation and their bearing on the participant's trouble, thus firmly casting "we" as an organisational constraint for solving that trouble ("we haven't got [...] the resource", lines 45-48).

The final usability report from site A2's testing includes the following findings: the "issue" that the "purpose and benefit of logging in or registering an account in order to complain was not clear" and a corresponding "recommendation" to "strongly consider whether it is necessary for consumers to register an account to make a complaint". Tracing the implication of these findings forwards, the completed live site for the media complaints organisation was redesigned to include an optional registration process; thus complaints can be submitted *without* an account, while retaining key account features for other users of the site, such as trade bodies or advertisers requesting assessments of their copywritten text. In that sense the finding being produced and worked on in this fragment seems to have led to formal reporting and subsequent impact on the resulting implementations of design revisions.

Stepping back both from this fragment, we can see that usability lab work practices themselves (and how they unfold and proceed) are deeply bound up in the production of the test's collection of distinct findings, which then of course may be taken forward to formal reporting. We can also say that it is apparent there are ever-present client-organisational matters incipiently at play [84, 91] in the constitution of such lab work; that is, they loom in the background, there as ready resources to turn to as members of the setting deem it relevant. Speaking to this point, we saw how the personnel 'bandwidth' ("the resource") of the organisation's complaint-handling process was brought to bear on shaping the way in which Adam accounted for P6's heard irritation about being presented with a login form during the task.

**Summary.** In this fragment we have seen how stakeholders are actively looking for 'trouble', working this up moment-by-moment and establishing trouble's relevancy. Importantly, there can be a distinction between *participant's trouble* and *stakeholders' version of trouble*—in this case contrasting the problem of being a "regular user" (participant) as compared with users "acting as police" (stakeholders). Here this turned on invocation of the client organisation, its human resources, and organisational procedures.

The next section provides a negative example of this phenomenon: where trouble can be made to 'go away'.

## 4.3   Dissipating troubles

As a matter of the test's work it is also the case that troubles and solutions *do not* necessarily lead anywhere. Ongoing work in determining relevancy forms part of the routine and familiar patina of usability testing as a work site. Of course, many seen-troubles may simply be ignored by OR members, who jointly but silently treat particular participant troubles as irrelevant for a variety of reasons that preclude their surfacing and eventual formulation as findings. Sometimes what might initially appear as prospectively relevant troubles, turn out not to be after discussion. Or, it might be that stakeholders' observation work deliberately involves surfacing something that is treated as non-relevant trouble from the start, as if to anticipate and account for what might otherwise have been joint silence. By examining one of these latter cases, this next fragment gives us a glimpse into the various contingencies at play.

I have suggested a couple of potential ways in which troubles seen to surface in the TR may begin to be formulated but still are ultimately 'passed over' by the OR and, with varying degrees of complexity, lead to what I am calling here a *dissipation* of the trouble (a reminder: that some troubles never make it to bone fide findings is the reason why 'trouble' and 'finding' have been distinguished). Like the formulation of troubles that lead to findings, it is out of the scope of this paper to exhaustively cover the different trajectories with which trouble dissipation might be reached, but, mirroring Section 4.1, here are a few example forms that employ different relevancy devices:

1) a trouble may be dissipated by topicalising the status of the object under test, for instance a prototype *as not an intermediate form*, e.g., that as a prototype the artefact will necessarily be incomplete or non-functional in parts, and that participant troubles might be seen to stem from the character of prototyping work itself;
2) participants might be seen to encounter a 'known issue', e.g., a trouble 'already seen before' several times by OR members and therefore may be acknowledged but not lead to further findings being formulated;
3) troubles may be dissipated as they are seen to be phenomena produced by the practical features of testing work;
4) the scope of 'what is being tested' or 'what the test is really about' can also be used to dissipate participant trouble.

**Fragment B ("s'a bit wireframey"): Description and transcript.** This fragment presents an instance of dissipation type (1) above. We join a different participant (P2) in the TR to whom the moderator (Helen, M) has recently asked "what was your overall impression of the website?". Responding to this, P2 has been offering various assessments of the site, during which she navigates to the home page, and then subsequently clicks onto the "Rulings" page (Figure 4). This presents a mocked-up list view of various rulings made by the client organisation about prior complaints that have been upheld or not. At this point P2 begins to produce various negative assessments of this page, and other pages she has just seen in the prototype, which is the point at which we join the transcript below.

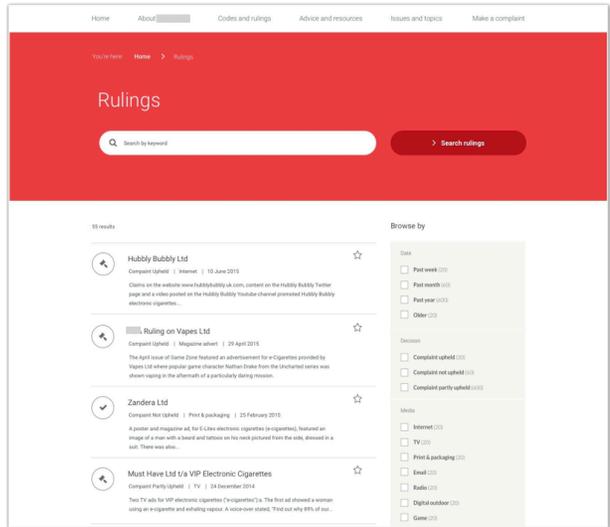

Figure 4: The "Rulings" page

| TEST ROOM (TR) | | OBSERVATION ROOM (OR) |
|---|---|---|
| P2  umm (0.3) I would say it's qui::te like with- what pages (w's) I just on then | 01 02 03 | |
| M   um: just now? [(           ) | 04 | |
| P2            [oh codes 'n stuff (1.1) | 05 06 | |
| M   i's- (0.5) very um (.) wordy? (0.6) | 07 08 | |
| M   [right | 09 | |
| P2  [(and at times though on:: >what I was saying about the titles< and things I think there could be more use of (the) colour (0.5) and stuff in those (0.3) just because it's on your eyes it's then quite it's very white? (        ) | 10 11 12 13 14 15 16 | Tom   °s'jus'poor IA° Tom   ((turning to Adam)) s'a bit |
| | 17 18 | wireframey this I think (though isn't it) |
| <REMAINING TALK IN TR OMITTED> | 19 20 | Tom   [still needs a bit of (.) bit of love= |
| | 21 | Jack  [yeah |
| | 22 | Adam  [yeah |
| | 23 | Mike  [yeah |
| | 24 | Pete       =(it was pretty quick)(    ) |
| | 25 26 | Jack  don't know if we (   ) (design so much) |
| | 27 | Adam  but the (.) yeah. |

In this fragment I want to focus on the ways in which this particular participant's assessments at this moment are heard by members of the OR as accounts of trouble, but treated and dealt with in a very different way to the previous fragment (viz. dissipating them and rendering them non-relevant for all practical purposes). In the TR, P2's assessment of the prototype as "wordy" leads to her referring to a prior moment where she was discussing the visibility of headings on the main pages ("what I was saying about the titles"). But, as she says this, over in the OR, Tom turns his head downwards, looking at his hand and quietly says "s'jus'poor IA" (see inset figure). This action is positioned by Tom almost as a (self-) criticism of the design of the information architecture (IA) elements of the prototype, gaining its sense from P2's negative use of "wordy" and reference to IA-relevant categories such as "titles". Tom's initial move here—like the previous fragment's group laughter and glancing—is also somewhat tentative, performed as 'self-talk', quietly. His prefacing "just" acts as a way of diminishing the significance of the trouble. Further, Tom's utterance, similar to Fragment A, rearticulates and transforms P2's stated trouble. Back in the TR, P2 continues with a more positive 'suggestion' that "there could be more use of (the) colour". However P2 then contrasts this immediately with a further assessment, "because it's on your eyes it's [...] it's very white?"; this last assessment is treated by the OR as a further criticism in need of an account. Hence, at this very point (line 16) Tom turns to Adam and provides his own, alternative (negative) assessment of the page: "s'a bit wireframey this". Tom continues to build on this himself, stating that the prototype "still needs a bit of [...] love". Here Mike, Adam and Jack all produce an overlapped "yeah" in apparent agreement with Tom's current and prior assessments (lines 21-23). Pete, the other designer present, then follows Tom's utterance with a comment that the work done on "it" (the prototype) "was pretty quick". After a murmured turn from Jack, Adam initiates what seems like an attempt to shift topics ("but the"), however ultimately pauses and turns to agreement once again (line 27).

**Fragment B: Discussion.** We might summarise what is happening thus: in the TR, P2 is articulating certain kinds of trouble, producing various assessments of the pages she has seen previously or has in front of her, while alongside this, members of the OR are moving to account for what they hear as that trouble in ways significantly dissimilar to Fragment A. Critically, there seems to be no solution being worked up here in the OR. There is nothing that gives us a trajectory towards a formulated usability finding. Instead, members of the OR work to dissipate that trouble by collaboratively establishing a sense of the provisionality of the prototype. They do this by transforming P2's troubles with "wordy" and "white" pages to various accounts *of* the prototype itself and prototyp*ing* as an activity. By highlighting the features of 'normal' prototyping work itself—that prototypes have shortcomings, that they are incomplete, that they do not have significant time spent on them, and so on—the members of the OR are able to build a characterisation of TR events and thus provide for the possibility of passing over this candidate trouble. (We can say 'candidate' since clearly at least some members of the OR considered *something* in need of accounting for: i.e., they are pointing out just why *these* witnessable troubles P2 is articulating as problems with the prototype are actually, really non-candidates and non-relevant.)

The work of dissipating emerging trouble—in tamping it down just as it bubbles up as a seen-possibility of a finding—involves OR members categorising activities in the TR as this-or-that *without* blaming the participant, or the moderator. Instead they move to diminish the relevancy of what it is they jointly agree they are observing, and they do so in this case by 'blaming' the artefact under test, i.e., the prototype, its status as a prototype, and all the limitations that might entail. The standard disclaimers included in the introduction for all participants is important here. This involves describing for participants how they are not the subject of the test (the moderator states "we are testing the website design, not you, so nothing you do or say is wrong") as well as the kind of stance the moderator will take during the test ("I have no personal involvement in the design of this site, so don't be afraid to say what's on your mind"). It is important to note that disclaimers such as these not only institutionally frame the unfolding interactions between participant and moderator in the TR, but also (repeatedly) frame *how* such TR interactions might be dealt with in the OR. In other words, troubles that the participant is seen to encounter during the test come to be seen in the OR as "normal, natural troubles" (to use Garfinkel's phrase [33 p. 187]) of the kind that 'anyone' might encounter when using something that is a prototype, and thus not to be blamed on this or that specific participant. OR activities are accountable to this established moral order of blame and in doing so enable the work of dissipation.

**Summary.** While Fragment A introduced the ways in which stakeholders *look* for trouble, this fragment examined how such trouble can also be made to *go away* (dissipate). There are different ways this might happen. Here, stakeholders leveraged the familiar features of prototyping work (wireframing).

Next we turn to problematising this (currently binary) distinction between bone fide findings and candidates that are dissipated.

### 4.4 Unpacking the contingent character of formulation work

What we have seen in the first two fragments is members of the OR sorting the relevant from the non-relevant: asking themselves, is this trouble a possible finding, or is it a candidate for dissipation? But the relatively straightforward character of these two fragments belies the complexity of the contingencies that *might* be brought to bear on determinations of relevancy and how the line between dissipation or eventual selection as the basis for a finding becomes blurry (see [19] on "possible" and "probable" findings). Accordingly, this section presents two rather more complex fragments; they deepen and problematise the present analysis of how trouble is surfaced in the OR, and unpick the tricky relationship between trouble and solution formulation. Overall they attempt to further underscore the strongly contingent character of formulation work. Once again, the purpose of this paper is *not* to assess the reasonableness or otherwise of judgements made by practitioners about what constitutes a moment bound for dissipation, and what does not, nor to provide an evaluation of the correctness of the various devices in play to do so (such as 'artificiality' / 'realism' or the nature of prototyping); rather, it is to offer descriptions of how stakeholders, as members of the test's work site, put forward and treat those judgements.

**Fragment C ("sometimes a bit artificial in user testing"): Description and transcript.** Dissipation is potentially a controversial and unsettling feature of usability testing if we adopt the view that findings are readily available intrinsically *in* artefacts and testing's job is merely to somewhat mechanically extract them. To push this further I now want to examine a moment of dissipation that is not informed by any participant account of trouble (unlike prior fragments) and instead underlines how troubles are 'seen' emerging. This fragment offers an instance of (3) as described in Section 4.3, i.e., that *troubles may be dissipated as they are seen to be phenomena produced by the practical features of testing work*. Fragment C here thus shows that trouble is *not* always dissipated in a way oriented by explicit accounts of participant trouble as in Fragment B but can also be dissipated via a 'behavioural' analysis by stakeholders. But it also brings into question any clear *binary* distinction between troubles bound for relevancy or non-relevancy.

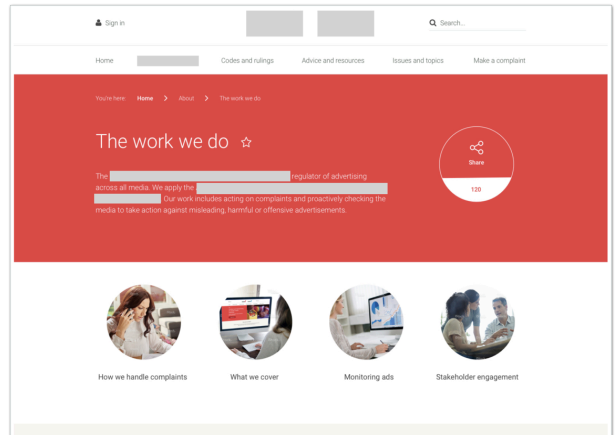

Figure 5: "work we do" page with the "four icons" mentioned by Tom (bottom)

A couple of minutes after the previous fragment, the same participant, P2, has now returned to the homepage (Figure 5) and is talking about how she would locate certain information on the site. The transcript below begins while she is doing this, with Tom observing to Adam that P2 "didn't scroll down below the four icons", likely referring to the bottom elements of the page depicted in Figure 5. (Note that the TR is *not* shown in the transcript below.)

```
                          OBSERVATION ROOM (OR)
01    Tom    ((turning to Adam)) she didn't scroll down below the four icons did she
02           (0.3) at all
03    Adam   No
04           (0.6)
05    Tom    she just stopped every time
06           (0.5)
07    Adam   no scrolling (1.9) >could you<, (0.5) >could you do< (0.5) what you did with
08           (        ) arrows >(I mean) s'thing< t- kind of show (0.3) something down.
09    Tom    (        )
10    Adam   >(bu- then)< that (.) I guess that- (0.5) adds stuff to the page a bit
11    Tom    's (0.6) s'just workin on the spacing below the (.) (panel's) j's quite a
12           lot (     ) befo::re (0.3) you can see: anything (.) else (on there)
13           (2.3)
14    Mike   I think it's sometimes a bit artificial in user testing and in reality they
15           would scroll down [     because they're here the- sort of
16    Adam                      [mmm                          ah, oh yeah=
17    Tom                              [   yeah right
18    Mike   =we're very focussed on doing the task
19    Adam   yeh
```

Before we get to the move to dissipate in this fragment, the first thing to note is that unlike all other fragments thus far, stakeholders in the OR have not surfaced any participant account of trouble articulated via TR talk. Instead, OR members spot a different kind of trouble phenomenon: an absence—a participant *not* doing something. Hence, Tom's opening assessment directed at Adam initiates things with an account of the trouble: that the participant "didn't scroll down". This account is tagged with a questioning intonation that leads to agreement from Adam ("no scrolling"). With the trouble surfaced and jointly acknowledged, Adam then moves to produce a candidate solution of using "arrows [...] t- kind of show something down". This is then is mitigated somewhat by him with the acknowledgement "I guess that [...] adds stuff to the page". In the next turn-at-talk this solution is then further developed by Tom who suggests that it actually involves "just workin on the spacing" that is used to arrange elements of the page. In many senses, this is very similar to formulating a finding, except of course that here the OR works to articulate the trouble *first* rather than solution first. It is possible that this ordering is coordinated with respect to the fact that any verbal articulations of trouble in the TR are not available to stakeholders here, so that trouble needs a different kind of formulation work to surface. In contrast, prior fragments have shown the participant *themselves* providing a first characterisation of trouble as part of the think-aloud which then gets picked up and transformed by stakeholders.

This said, and returning to the topic of dissipation, the most critical feature in this fragment is Mike's interjection on line 14, which offers a move to dissipate Tom and Adam's present trouble-articulation and iterative solution-work. Mike does this in such a way as to recategorise the observed trouble—the scrolling and what the stakeholders see as its absence in P2's conduct—with an account, instead, of how P2 would be acting "in reality": "I think it's sometimes a bit artificial in user testing and in reality they would scroll down". Mike then continues to elaborate this recategorisation, interspersed with agreement from Adam and Tom. Mike builds on this in the following way: that "because they're here"—'any' participant in 'any' usability testing lab—it is *the task* that is the priority for the participant ("we're very focussed on doing the task") and thus in competition with 'realism'.

**Fragment C: Discussion.** In producing usability findings, the act of seeing trouble in TR activities as they are made available audio-visually in the OR is itself an (ordinary, routine) accomplishment by members of the OR. As we have already discussed, trouble must be searched for and actively seen by the OR, and through this rendered available for co-present others. It is precisely in doing this that it then becomes possible for members of the OR to 'dispute' those ways of seeing trouble; in other words, seen-troubles allow for different moves to be made with them in OR. Here, the relevancy of the source trouble (scrolling down and missing the rest of the webpage) is dissipated behaviourally by reference to the nature of testing itself and an appeal to 'what participants do' in testing compared with what 'anyone' does in 'normal' settings. The method of dissipation here includes Mike's following assertions: that usability testing is "sometimes a bit artificial", that it contrasts with "reality", that normal behaviour would be to "scroll down", and that the test focusses on "doing the task", i.e., that usability testing is knowingly performed ironically.

As the most experienced consultant there is a sense in which Mike—perhaps in a similar way to the next fragment—is doing *tutorialising* work here for the less experienced stakeholders, as well as establishing what it means to appropriately 'see' (not just ocularly) and discern certain kinds of TR happenings as relevant, and others as not. In this particular moment we might be tempted to say that Mike is co-orienting with other stakeholders to distinguish relevance for different domains of action: one called "user testing" and the other called "reality", and then by implication using this as a categorisation of *this* participant's visible activities and the 'right' way to see them. He does this as a kind of teachable moment for the others in the OR, who, it is worth noting, are still relatively new to this test having only seen P1 previously (and a subset of them not having experienced usability testing at all).

In contrast with Fragment B, it is not clear whether Mike's intervention is necessarily an attempt to *completely* dissipate the ongoing exploration of possible design solutions Tom and Adam are working on here. However, Mike's interjection does close down this particular sequence within the OR by shifting topics from 'solving this trouble' to 'the nature of tests', which (after this fragment ends) leads to the OR returning to silence ready for the next observation to be made.

It is important to point out that this fragment suggests a more nuanced notion of dissipation. During testing and the surfacing of trouble, there are often various unstated adjacent troubles that are connected with the current one. In this case, there has been—and will be after this particular participant—a recurrent discussion in the OR about the size of the banner on the home page (for visual reference, see the red banner visible in Figure 5). The particular specific trouble in this fragment around a lack of scrolling is indexed to this backdrop of related troubles. Although there is a move to dissipate here, in the final usability report there is the following issue noted: "Due to the size of the image banner, some participants did not know that they could scroll down beyond this. As a result, they did not see any more of the homepage than was initially visible." Ultimately in the final website the banner *is* lessened in height from the prototype version by about 0.75cm (i.e., ~90% of the prototype's banner), in line with the report's recommendation to reduce it. Thus the trouble in this fragment ultimately ends up being brought together with the other problems that centre around the banner size. The point is that although an explicit connection is left unarticulated here, this fragment's trouble is nonetheless part of a broader contexture of 'troubles like this one' that exist across and connect multiple sessions and participants.

This all suggests a more complex relationship between what is ultimately considered relevant and non-relevant in findings' production work. Here, there is what looks like a possible move to dissipate by Mike, but which likely serves more significantly as tutorialising moment: perhaps acting as a joint 'warning' and instruction about 'how to see', rather than a strong attempt to delete this particular trouble from play entirely. In other words, dissipation is not always a straightforward binary matter.

**Summary.** Stakeholders need to develop *critical* ways of seeing trouble and its dissipation that continuously consider the situation—the 'testness of testing'. This fragment has explored what counts as 'normal' participant behaviour in a test, and how troubles that emerge might fit alongside this ever-present concern. Stakeholders are working to find *the sense of the trouble* for the test circumstances they are in.

Next we look at how troubles are connected with a critical part of testing: that of finding potential *solutions* to the trouble.

**Fragment D ("interesting"): Description and transcript.** This final fragment is perhaps the most complex, allowing us a much closer look at the contingent features of findings' formulation work and its relationship with dissipation, as well as troubles and solutions. Here the moderator (M, John) has just talked the current participant (P1) through the standard disclaimers that preface the very first task. The first task for P1, as introduced by M, is to "make a complaint" using the prototype website's "Make a complaint" page that is visible on the screen in front of P1 (Figure 6). In the scenario, P1 has been told that there is a picture of the advert they wish to complain about on the computer's filesystem, which they will upload during the complaints process. Sitting in the TR with the prototype in front of them, M asks P1 "if you can explain to me how you think you'd go about doing that" (i.e., "make a complaint"). This then leads to P1 reading out loud some of the text visible in Figure 6, particularly the section marked

**Figure 6: The "Make a complaint" form and "Continue" button (bottom); section P1 is reading has been marked**

"What you will need".

| | **TEST ROOM (TR)** | | | **OBSERVATION ROOM (OR)** |
|---|---|---|---|---|
| **P1** | "two to three minutes to provide your details" (0.6) uh (3.0) okay (.) yeah so I'd erm I would (.) email (0.3) the [media complaints organisation] and I'd um (1.4) attach (0.7) a ph<u>o</u>to, (0.6) | 01 02 03 04 05 06 07 | **Jack** | interesting |
| **M** | s- so you would (.) you'd <u>send</u> an email | 08 09 | **Pete** | ((glances towards Tom with a grimace)) |
| **P1** | <u>yea</u>h ye[h | 10 | | |
| **M** | (         ) [so th- how'd you think you'd do that (0.6) w- where'd you think you'd send that to then (0.6) | 11 12 13 14 | | |
| **P1** | w<u>e</u>ll (.) erm ((looks about screen)) (3.0) (('scanning' screen)) | 15 16 | **Tom** | °(need) to make it clearer° (.) (the button) |
| **P1** | °(   )° well it doesn't s- it doesn't say anything about contacting (us on here) <br><br> **<REMAINING TALK IN TR OMITTED>** | 17 18 19 20 21 22 23 | **Jack** **Mike** | it's interesting that ((towards Helen)) but then (that tha-) if she someone says they'll send an email just say that's that's a <u>brill</u>iant way of doing that |
| | | 24 | | (0.7) |
| | | 25 | **Helen** | yeah |
| | | 26 27 | **Mike** | per<u>haps</u> let's try this f(h)orm ((glances towards Tom)) |
| | | 28 | | (0.7) |
| | | 29 | **Helen** | okay |
| | | 30 31 32 | **Tom** | ((turning to Mike)) <u>th</u>at's er- we can make that button clearer (.) right?= |
| | | 33 | **Mike** | =yeah= |
| | ((P1 clicks home button)) | 34 35 | **Tom** | =c- make a complaint (0.3) instead of continue |
| | | 36 | **Mike** | yeah |
| | | 37 | **Adam** | (aye) |
| | | 38 | **Tom** | yeah |
| | | 39 | **Mike** | yeh |
| | | 40 | | (1.6) |
| | | 41 | **Tom** | ((on laptop)) °do that° |

Since what is happening within this transcript is quite complex and involves understanding pairwise activities between the TR and OR, I first want to go over what unfolds, moment-by-moment in the various adjacent (but notably asymmetric) relations between the occupants of the two rooms. First, we should consider the following moment:

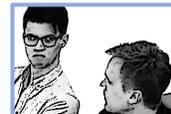

| **P1** | I'd um (1.4) attach (0.7) a ph<u>o</u>to, (0.6) | 05 06 07 | **Jack** | interesting |
|---|---|---|---|---|
| **M** | s- so you would (.) you'd <u>send</u> an email | 08 09 | **Pete** | ((glances towards Tom with a grimace)) |

P1 (belatedly) responds to M's question (not transcribed) with "I'd um [...] attach [...] a photo". This occasions Jack, over in the OR, to utter a vague assessment, "interesting"; this utterance comes to be placed sequentially in the 0.6s pause between P1 and M's utterances. Jack's temporal placement of this utterance here means that "interesting" can be heard as a local response to P1's answer to M about emailing an attached photo to the client's organisation. In other words, this remark by P1, coupled with what is visible on-screen (Figure 6, itself an account of P1's progress through the task), is made relevantly noticeable to others in the room *by* the production of Jack's "interesting". Once again, this is surfacing work by Jack.

Meanwhile in the TR, M confirms this response from P1 with a question about the hypothetical email: "so you would you'd send an email [...] where'd you think you'd send that to then". M's job here is in balancing the need to keep the participant on-task but at the same time 'let them lead' during the test. During this, Pete glances towards Tom and pulls a face (see inset figure above) which, like Jack's "interesting", gains its significance from its temporal and sequential placement and suggests an attempt to initiate further actions from co-present others [39]. In this case Pete's gaze and expression, building on Jack's "interesting", thus acts as yet another (bodily) proposal for the relevancy of what P1 is doing, and in turn the relevancy of that to the OR's collective interest in locating usability findings. This is much like the scattered laughter and glances of Fragment A.

Meanwhile, things have moved on in the TR; P1 responds to a question from M about where to send the (hypothetical) complaint email to:

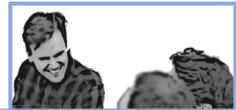

```
P1    well (.) erm ((looks about screen))      15
      (3.0) (('scanning' screen))              16    Tom    °(need) to make it clearer° (.)
P1    °(   )° well it doesn't s- it doesn't    17           (the button)
      say anything about contacting (us        18
      on here)                                 19    Jack   it's interesting that
```

P1 appears to be searching the screen in front of her, still on the "Make a complaint" page as per Figure 6 (later she will click the "home" button on the page, thus leaving the "Make a complaint" page). During P1's turn to visual searching, Tom treats this moment as an opportunity to (quietly) offer a first attempt at a solution for the assembled members of the OR: "(need) to make it clearer (the button)" (see inset figure above). This gnomic utterance from Tom is sequentially positioned in such a way that it acts as a response to number of proximate actions by others both in the OR and TR: Jack's "interesting", Pete's glance and mild grimace, as well as P1's ongoing search for 'what to do next'. In other words, these priors provide a staging post of sorts for establishing the sense and relevance of Tom's utterance *as* a possible 'solution'. Tom also produces this solution after looking downwards, away from the P1's projected screen, scratching his head and then turning towards Jack in particular with grimace that turns into a smile, perhaps, we might say, of trouble recognition (inset figure). Jack then repeats his earlier assessment ("it's interesting that") marking a momentary discontiguity between Jack's project—which is remaining non-specific about the trouble—and Tom's project—which is to reach a solution to the seen-trouble immediately via his a candidate analysis of P1's apparent searching as she turns to the screen. During this time, P1 begins to articulate her trouble explicitly with "it doesn't say anything about contact us on here", thus offering another resource for the OR to build its shaping up of the solution (and trouble).

However, at this point Mike, turning towards Helen, gives some advice to her about moderating further tests (note Helen is a trainee moderator, and she will be performing the next session). This acts as a move *away* from what Tom, Jack and Pete are doing. During this, Mike glances at Tom, widening the relevance of this advice-giving to the rest of the OR. Subsequently Tom turns away from the screen and instead towards

```
30    Tom     ((turning to Mike)) that's er-
31            we can make that button clearer
32            (.) right?=
33    Mike    =yeah=
34    Tom     =c- make a complaint (0.3)
35            instead of continue
36    Mike    yeah
37    Adam    (aye)
38    Tom     yeah
39    Mike    yeh
40            (1.6)
41    Tom     ((on laptop)) °do that°
```

Mike, *revising* his prior solution (to make "it" clearer) into a search for agreement from Mike about this proposal, saying "we can make that button clearer (.) right?", while at the same time underscoring a new sense of the provisionality of the solution by appending a questioning "right" (the "button" referred to is the "Continue" visible in Figure 6). Tom's question appears to be sensitive to Mike's prior moves. Tom's appeal seems to work since Mike then quickly agrees, and Tom further elaborates what was previous left unclear (making the button clearer, but how?); specifically this is a more developed solution that modifies the label of the button to "make a complaint, instead of continue". After further agreements in turn from Mike and Adam, Tom turns to his laptop, quietly saying "do that", and begins typing.

**Fragment D: Discussion.** The OR's production work here mostly foregoes any explicit articulation of trouble that we have seen in prior fragments and instead moves rapidly toward a solution. In this case the seen-trouble indexes P1's response to the task in various ways (her response being to send an email rather than use the web form in front of her). Even moreso than Fragment A, there is a momentary reversal in that stakeholders begin with the solution and in this case barely articulate what the actual trouble is. To begin with, it is important to consider that this is the first test participant. Mike intervenes with a 'teachable moment' (somewhat like Fragment C) by addressing Helen in a way that acts as a indirect counterpoint to the potential 'rush to the solution' being built in the unfolding interactions between the designers. Like Fragment C, such moments do not necessarily end in a simple dissipation but rather offer particular orientational resources for other stakeholders as the test proceeds.

That said, this fragment also pushes the complexity of the relation between trouble and solution even further and offers a vantage point for more reflection on this. It appears counter to an HCI literature that tends to construct clear daylight between the trouble and solution and also one that typically presents a clear preference ordering (problem identification → design solution). The point is that while trouble *is* often articulated by the members of the OR, sometimes this may not be the case. Design solutions do not always need formally stated articulations of usability trouble as their warrant in testing. Instead, troubles may remain largely unstated and only gain their sense as findings later. This further deepens observations from Fragment C where the 'contexture' of troubles established across sessions. For instance, in this case there *is* a formally-reported finding in the usability report that seems derived from the implied as trouble here. But it acts as an amalgam and perhaps transformation of this and (crucially) other subsequently observed troubles: "The process of clicking *Continue* after every few fields was cumbersome as it added superfluous interaction after each field or group of fields had been entered." Naturally, a solution (labelled "Recommendation") is also present: "The final button should not read *Continue*. Instead, it should read *Submit complaint* or similar." Notably, the actual design for the completed live website for the client organisation uses "Back" and "Continue" buttons during the complaints process, with the final stage offering a "Submit Complaint" button.

I now want to make a few more detailed observations about trouble and solution formulation and its collaborative characteristics that build on and go further than my comments on prior fragments (particularly Fragment A). Jack's initial move here—uttering "interesting"—is a minimalist gloss. It is specifically non-specific at this stage, but, through its sequential and temporal placement, points to and offers a topicalisation in the OR for what is seen happening at that moment in the TR. In a sense it acts as a 'call to attention' and a way of cohering or 'gearing into' the work of findings' production. It starts to surface and make available for other OR members the prospective relevancy of P1's actions *without* articulating what the trouble actually is. Through this vagueness, Jack offers what seems to have been observed as a contestable object. This serves another pragmatic purpose. Members of the OR are of course *not* continually watching the test unfold, and instead tend to 'dip in' and 'dip out', not only attentionally but perhaps also physically (e.g., as was the case for Mike, who had other competing duties to attend to throughout the testing). The problem is that there are a great many other concurrent activities going on during the test, such as getting the next participant ready, writing notes, taking toilet breaks, getting refreshments, attending to email and other work activities, talking about findings (and whether they 'really are' findings) and possible solutions from prior participants' sessions, and so on.

Further, the vagueness of Jack's "interesting" does not close off the identification of trouble sources seen in P1's actions and instead creates latitude for other OR members to themselves produce further analysis. We can thus see in this fragment how the OR collaboratively and developingly works-up this somewhat 'bare' utterance sequentially into an increasingly elaborated solution (and its implied trouble). Tom's initial solution is at first to "make it clearer". While it is not clear on what "it" might be, this is then immediately narrowed within his turn using an appended mention of "the button" (line 17). However, there are *many* candidate buttons on the page in question: so, Tom is relying upon a shared seeing of trouble in question. Tom later reiterates this in his response to Mike (line 31), emphasising the relevance of "that button" (*which* button still being implicit): "we can make that button clearer". It is only in the final stage of this progressive development of a solution (which Tom further elaborates on lines 34-35), where Tom provides explicit reference to the button in question (i.e., the button with "continue" as its label), and its solution, to change it to "make a complaint, instead of continue". To summarise, what we see here is that the gradual progressivity of this formulation work in the OR, dealing with troubles and solutions, is very much a collaborative achievement.

Finally, how to treat the implicit trouble is *not* roundly agreed upon immediately in the OR, for we see Mike orienting to it with a different response (a move away, to dissipate it) alongside Tom's suggested solution. The move here by Mike offers one way to potentially dissipate this particular seen-trouble by treating it as not in the *scope* of the usability test. To recall, this is an instance of type (4) in Section 4.3: *the scope of 'what is being tested' or 'what the test is really about' can also be used to dissipate participant trouble.* Mike's move here is seemingly not taken up by others: subsequently Tom reiterates his original candidate design solution in response to Mike's turn and as part of this seeks agreement from him, i.e., "we can make that button clearer right?". Mike offers no resistance and instead provides an agreement and thus in a sense the potential dissipation itself is dissolved. But there has been a measure of interactional friction in the course of getting to that agreed-upon solution (and implicitly its trouble). This friction is produced as Mike topicalises the trouble as relevant for test and moderation practices *rather* than it giving warrant for a solution. He initially designs the advice for Helen, advice which indexes things like developing moderator skills to 'lead without leading', 'manage participants', and establish a normative notion 'what counts' as within the scope of the test. But Mike's tutorial is then expanded to the room through his glance towards Tom. It is no coincidence that such teachable moments are more prevalent within the earlier sessions: firstly to counter the potential early focus on design solutions before the articulation of trouble; and secondly to jointly establish situationally specific, appropriate ways of actually seeing trouble within the observation room.

**Summary.** A distinction must be made: there are troubles but there are also *solutions* being worked up. The ordering of these can vary. Trouble may be surfaced first, then the solution (Fragment C "no scrolling" → "spacing"). The solution can be first, then the trouble (Fragment A "we need to get rid of that" → "we haven't got the resource"). Or, in this case, we might go straight to a solution ("make a complaint, instead of continue"), in which case trouble will need to be worked up later in report (deferred).

## 5 DISCUSSION

As a methodic achievement, stakeholders involved in usability testing collaboratively work towards producing findings and treat one anothers' actions as similarly oriented to that end (and hold them to account should there be a breach of this). Even a laugh or a turn of the head can initiate and ultimately work towards a finding being shaped for all present into an "increasingly definite thing" [35]—a social object that something must be done with.

In detailing Fragments A-D, this paper sheds light on the *incipiency* of finding-formulation and the *contingencies* at play in a lab's usability testing work. What is incipient here for stakeholders is the possibility of seeing trouble and bringing to bear its significance for the test. And, as we have seen, there are many different types of trouble that stakeholders continuously and motivatedly do seek to discover in the course of test participants' accomplishment of a set of tasks. Importantly, what the stakeholders are discovering, though, is trouble 'on their terms'. They do not slavishly adopt participant accounts of trouble, however, but instead appropriate those accounts as interactional resources in their deliberations. In a sense they are performing (in this case) a kind of "vernacular video analysis" [65] of the live audiovisual stream broadcast into the OR. As a reminder, in Fragment A the participant reacts badly to the requirement to create an account during the complaints process. But the observation room's version of trouble is markedly different from the participant's in that it becomes mitigated by organisational needs and other types of user. This element of transformation was also present to a varying degree in the other fragments. In other words, there is a *generative* rather than linear quality to the way the TR proceedings shape how members of the OR look for trouble [40].

Trouble *relevancy* here is also key. Stakeholders in OR are not merely 'detecting' a signal present in the test and neither are they merely 'reading off' troubles. Instead they are *motivated observers* in that they are persistently (and collaboratively) looking for, surfacing, and shaping up trouble as matters contingent upon a raft of relevancy devices: e.g., is it a finding, or just a feature of how the test has been configured? is it a finding, or just a property of prototyping? is it a finding, or is it a problem with the test and how it is being run? At all times establishing this relevancy is critical. Sometimes these collaborative, progressive formulation practices lead to 'bone fide' or 'ratified' findings, while at other times relevancy devices are deployed to bring the status of candidates into question in some way and thus dissipate them. At other times matters are not so clear given the complex implicative relationships potentially held between different seen-troubles across sessions. Ultimately, only some troubles end up reported as findings, in part because findings are contestable and *produced so as to be contestable* (see [8]). This is critical since contestability and what might count as contestable is itself shaped by relevancy devices embedded in the situated and contingent arrangement of testing-in-practice: this particular artefact, this particular set of tasks, this particular selection of participants, serving this particular clients' interests, these specific stakeholders in the room, and so on.

In another context this observation of contingent 'sorting' practices would be nothing new. For instance, we might turn to Benson and Hughes' discussion of the work of interviews as a way to construct datasets for variable analysis in the social sciences:

> "In effect, the rules of interviewing are practical procedures for managing a social encounter in order to get the interviewing done; all focussed, we might say, to achieve meaning equivalence in the material. Thus, the interviewer, and later the coder, out of all the interchanges that take place on the occasion of the interview, have to select which are 'chatter', make judgements about the consistency of replies, and so on; all the myriad of practical decisions, judgements and interpretations that have to be made to get the material to speak in the way required. [...] In doing their best to do the work in accordance with the rules, interviewers and coders have to 'resolve ambiguities', 'let certain remarks pass', 'allow propriety to constrain lines of questioning', 'hold meanings in reserve', and more; in some, use their common-sense knowledge of social structures to make sense of the replies, the coding task and, later, make sense of the tables (Cicourel, 1973)." [4 p. 122]

In short, stakeholders moderating and observing usability tests face somewhat similar challenges to the interviewers and coders that Benson and Hughes describe (although the aims are necessarily different, of course). So, although this is not 'news' in a general sense, the particular challenges of usability testing and evaluation remain unexplicated.

There are significant conceptual challenges posed by this that call into question broader notions in usability of problem identification 'reliability'—a concept deeply baked into arguments about the number of test participants [64] and comparative evaluations [69, 70]—and whether it is indeed the right kind concept to use when considering how usability findings relate to the proceedings of the test. Thus, when one looks at actual instances of usability testing, the formulation of findings turns on: a) emergence from seen-trouble; b) their progressive realisation; and c) situational contingencies that may lead to findings' ratification, dissipation, or something in between.[8] Although I have focussed on the work of practitioners in an industrial setting, I would argue that this point likely extends to usability testing and its variants being put to use for HCI research (albeit with a range of *different* contingent concerns and relevancy devices at play). The alternative account offered here instead suggests practitioners performing something akin to (but not the same as) "discovering work" [89], in which the social "shaping" of the object (finding) is an "increasingly definite thing" [35]. This poses a conceptual challenge to positivist renderings of usability testing, and perhaps usability evaluation too—renderings that remain key to the intellectual endeavours of much HCI research on usability testing and its attitudes and orientations towards practitioner adoptions of such methods.

Finally, there is also a significant monetary cost and professional significance associated with running usability testing and delivering its outcomes, all of which plays into stakeholders' drive to produce findings during tests. A critical part of this work involves establishing an agreed order and organisation so as to manage the sometimes conflicting priorities and concerns of different OR members (designers, usability lab staff, clients, etc.)—a fact that is perhaps just as important to usability testing in practice as it is about producing findings.

### 5.1 Contributions to HCI

The first contribution of this paper is in providing an initial **unpacking of the internals of usability testing**, which, it seems, may well help begin to address some key conceptual problems in HCI's approach to usability testing. As Følstad, Law and Hornbæk have argued, "data from usability testing are challenging to interpret" [29], a statement that somewhat belies the complexity of what is going on when we take a close look at usability testing's actual practices. More often than not, usability testing is treated by HCI research as an activity that involves 'cranking the handle' to churn out problems, whereas this paper's study has suggested a) that this is conceptually mistaken, and b) that UX practitioners' work is in various ways practically different to academic HCI's conceptions of what usability testing is [98, 61]. Much HCI research on usability testing focusses on its accuracy and efficiency, seeking to fix 'broken mechanics' and replace 'outdated parts' with better, more controllable ones (as surveyed at the start of this paper). I do not wish to suggest that this large, rich body of work seeking to improve and understand the limitations of usability testing is somehow incorrect or misguided or unproductive. Far from it. Rather, I am suggesting that the work of establishing these 'fixes' (often derived from various perceived deficiencies in practitioners' work) will be impeded if HCI remains largely unfamiliar with the mundane, practical features of industry practice. And, developing the kind of understanding I propose might then lead to a better position for HCI to more concretely evaluate this or that 'fix' and both its applicability for and fit to practitioners' work.

I have also sought to show the significance for HCI in starting to investigate in a detailed way the practical 'internals' of testing, and its "interactions as products of a machinery" [81]. This means that usability testing's practical work—the collaborative and contingent achievement of producing findings—is a significant but largely unexcavated driving force in usability testing and possibly usability evaluation at large. Looking at the internals reveals, for instance, that emerging troubles may well dissipate, a point that seems to potentially challenge conventional assumptions about usability testing itself. To bring this all together for HCI, I offer a summary of this and other internals as follows:

- at its broadest, stakeholders in the observation room collaboratively and progressively work to surface, formulate, and deal with emerging, candidate troubles so as to (prospectively) shape them into usability findings;
- trouble surfaced and formulated in the observation room is trouble-*that-is-seen* (i.e., seen-as this or that object by stakeholders), of which there are two further important features:
    a) what constitutes noticeable and relevant trouble is shaped by collected contingencies at play;
    b) observation room troubles may be informed by participant accounts of trouble yet are not identical to them, e.g., they involve reformulating (*transforming*) what participants report;
- production work often does not (initially at least) *causally* arrange the production of design *solutions* from the *troubles* that participants are seen to be encountering or accounting for (i.e., it is clear that members of the OR might articulate potential solutions *before* any discussion (if at all) of what 'the trouble' even is);
- practitioners might not *necessarily* treat troubles and their solutions as separate or *only* to be addressed at specific phases of projects; entangled formulations of troubles and solutions then need to be disentangled or 'reconstructed' during the formal reporting process;
- candidate troubles may be routinely dissipated in various ways outlined in this paper.

---

[8] There is no sense in which I am arguing that this represents 'the whole picture'. Future work must necessarily delve deeper into usability testing practices and their many variants in order to provide a more comprehensive view.

Beyond offering a description of usability testing and an argument for its reconceptualisation, what other contributions are here for HCI? The fragments presented in this paper are by their nature partial. As previously noted there is actually a very wide range of possible kinds of trouble to be observed in the broader video data set. The video data also offers exhibits of the ways in which findings may be generated, yet are *not* occasioned by trouble a participant is seen to have, e.g., a participant suggesting a new feature. Rather than attempt to broadly cover the diversity of production work (which must be investigated in future studies), I have instead adopted a strategy to attempt to drive an initial wedge into the interactional work of producing findings, and in so doing show *how* future work in HCI might continue to investigate usability evaluation practices at large.

In that sense, this aspect offers the second contribution of the paper: a **programmatic pointer** to the study of usability testing and perhaps usability evaluation in general. It is programmatic since this study itself offers an initial set of "directions along which to look"—to use Blumer's helpful phrase [5 p. 148]—when considering how future studies, encompassing the diversity of usability evaluation methods, *could* be approached. I am not necessarily advocating an ethnomethodological approach per se, but rather emphasising in principle the potential value to HCI in paying much closer attention to—picking apart in a serious way—the very details of evaluation techniques that in a formal sense HCI is extremely familiar with and yet at same time in a praxeological sense remains quite incurious (and perhaps ignorant) of. Ethnomethodological investigation clearly has much to offer here on those terms.

A third contribution to HCI's understanding of usability testing that might be derived from this paper is in elaborating the particulars of **usability testing work sites as *sites for design***. It is hardly a new observation that the relations between testing methods and design methods are somewhat blurry, even explicitly so in the case of RITE [68]. For instance, some adaptations of usability testing embed design solution search within test procedures via sketching [92] or eliciting redesign suggestions [28, 27] (or, somewhat differently, by synthesising re-enactments of crowdsourced usability problems with design work [37]). But it seems that research on usability testing has yet to elaborate how design activities may be endogenous to testing and actually intertwined closely with the production of findings themselves. In mixing troubles and solutions it becomes clear that design work may be deeply embedded by practitioners into the very processes of testing itself—even the more 'classic' forms of usability testing that I have expounded in this paper. We saw how design solutions were progressively, interactionally developed, and mutually arrived upon, and that such design solutions may themselves stand as articulations of observed troubles in the TR (i.e., 'usability problems') that nevertheless remain somewhat implicit until a later stage of formal reporting. Aside from *hints* that this may be the case in existing literature (e.g., Hornbæk's challenge of the usability evaluation dogma of "evaluation in isolation from design" [51]), there seems to be an absence of work explicating the opportunities of common-or-garden usability testing itself as a design site. In this way the study presented here offers an empirical complement to Hornbæk's suggestion:

> "the defect-identification view captures only part of the role that usability evaluation plays in software development. It does not, for example, capture getting design ideas, communicating insights to designers and developers, or experiencing things other than problems in a usability test" [50 p. 270]

That said, the mixing of troubles and solutions seen in the fragments here may not hold for less heterogenous groupings of stakeholders, or cases where product owners or designers are not present and instead receive usability reports and highlights reels.

Nevertheless, that design can be so tightly and interactionally intertwined with the production of usability findings is consequential in two ways. Firstly and perhaps most obviously, the shaping of design outcomes from test findings has a significant impact on a wide range of commercial products and services, and it is likely that the way in which this happens in industry settings differs substantially to academic implementations of usability testing. How it is that some *candidate* troubles dissipate while others do not strongly shapes the contours of those prospective impacts. It also calls into question the typical framing of usability testing as just another kind of experiment rather than a site in which practical design work unfolds quite literally moment-by-moment. Secondly, conceptualisations of design that attempt to account for how design functions (such as design research communities in HCI) may gain greater descriptive power from seeking new sites of design that may have been traditionally overlooked, such as industry usability evaluation practices.

The final contribution of this paper concerns both HCI research and UX practitioners themselves. By **painting a rich picture of practitioners and their work practices**, HCI researchers interested in operationalising their findings (e.g., see [20]) can gain more purchase on the texture of practice itself. As I argued in the introduction, non-specific uses of the term 'practitioner' can unhelpfully gloss a wide range of precise and particular ways in which different industry settings engage with an ecology of actual practices: user research practices, design practices, software development practices, product management practices, business modelling practices, marketing practices, and sales practices (to name but a few potentially relevant features of practitioners' work that become leveraged in the development of products and services). This study suggests a complexity to this and to practitioners' concomitant concerns—a complexity that is often absent from typical characterisations in HCI. For instance, fragments from this study show how practitioners seem to orient to and work quite directly with various conceptualisations of testing. These are often not too distant from concerns expressed in the academic literature on usability evaluation. This includes practitioners deploying notions of realism in testing (e.g., artificiality), something similar to 'ecological validity', displaying sensitivity towards how the character of prototyping meshes with the contingencies and institutionality of testing, through to articulating normative conceptualisations of expected participant behaviour and participants' 'representativeness'. Yet rather than treat such matters as possible moves towards theorising testing, practitioners deal with the practical consequences of them here-and-now.

Taking the details of practitioner work seriously in this way is part of adopting what one could call a 'human-centred approach' to UX and wider design practice. This is important if HCI intends to provide meaningful outputs that support and give assistance to practitioners in various ways. Equally, for practitioners themselves, detailed reflective accounts of practice such as this could form possible training aids that sit somewhere in between textbook learning and craft acquisition.

## 6   CONCLUSION

In this paper I have argued that usability evaluation, and usability testing specifically, are typically presented as processes in which findings are generated as a matter of course. In questioning this, the study presented here has closely examined the generation of findings as a production, as an achieved phenomenon. While much of the usability evaluation and usability testing literature has an understandable preoccupation with *methodological* issues (e.g., to what extent methods are reliable, repeatable, decomposable, etc.), this paper has instead attempted to reposition this by considering how usability testing is a *methodological achievement* by members of the setting (i.e., the stakeholders). The details of that achievement have been presented in the various ways those members of the test not only produce findings with their connected troubles and solutions, but also how findings may begin to be formulated but then dissipate instead.

In concluding, there are several lines of investigation that are implied by some of the limitations inherent in this paper and might further enrich HCI's understanding of usability evaluation and practitioners' work. I have mostly ignored how roles of different and sometimes competing or conflicting stakeholders are at play in testing, as well as how many broader organisational concerns [102] are enmeshed in usability testing and its work. The study also has done little to take into account in any deep way how findings evolve and mutate across participants and across sessions (and testing days); future work should address findings' production as part of a broader 'lifecycle'. Further, the key interactions between participant and moderator and the pragmatics of moderator talk including its design to enable progressivity of the test, have mostly been left unexamined. This includes any systematic inspection of how participant actions themselves come to be embedded in various membership categorisational forms [83] in different ways both in the test and observation rooms. This might be, for instance, how participants constructing themselves or others as 'this' or 'that' kind of participant, just as we saw with Fragment A where the participant positioned herself outside the category of what she termed "regular" users, with all its concomitant negative associations. This study also has only concerned itself with how a mid-fidelity prototype is tested, yet there are many other forms such as product baseline testing, or competitor product testing, all of which will impact how production work in both test rooms and observation rooms unfolds. The moments leading up to the test and after it have also been left mostly untouched, and studies should be conducted to examine task formulation (and "triangulation" [61]) as well as report generation. Finally, there are many other different test configurations that should also be considered, including remote testing, in situ testing, RITE, and guerrilla testing.

## ACKNOWLEDGMENTS

Permission has been granted by the copyright holder (the client organisation) to reproduce all screenshots of the prototype. This work is supported by the Engineering and Physical Sciences Research Council [grant numbers EP/M02315X/1, EP/K025848/1]. I would also like to thank the

following in (alphabetic order) for their helpful feedback on this work: Bob Anderson, Adam Banks, Elizabeth Buie, Teresa Castle-Green, Andy Crabtree, Joel Fischer, Christian Greiffenhagen, Neha Gupta, Carine Lallemand, Martin Porcheron, and Simone Stumpf. Several data sessions, held with members of Mixed Reality Lab and the Hong Kong-Macao Ethnomethodology and Conversation Analysis Group, have contributed significantly to the writing of this paper. Finally, I would also like to thank anonymous reviewers for their detailed and insightful observations on this work (and give apologies to anyone I have missed acknowledging). *Data access statement: Due to consent restrictions in place during the studies outlined in this paper including sensitivities towards protecting stakeholder identities and processes, only selected fragments of research data can be made available; please refer to http://dx.doi.org/10.17639/nott.375*